\documentclass{aa}  

\usepackage{graphicx}
\usepackage{multirow}
\usepackage{amsmath}
\usepackage{eucal}
\usepackage{gensymb}
\usepackage{orcidlink}
\usepackage{subfigure}
\usepackage{natbib}
\bibpunct{(}{)}{;}{a}{}{,} 
\usepackage{siunitx}
\usepackage{txfonts}

\begin{document} 

   \title{Towards direct imaging and orbital parameter estimation of supermassive black hole binaries with spaceborne VLBI}

   \author{B.~Hudson
          \inst{1}\fnmsep\thanks{\email{b.hudson@tudelft.nl}}\orcidlink{0000-0002-3368-1864}
          \and
          L.I.~Gurvits\inst{1}\fnmsep\inst{2}\orcidlink{0000-0002-0694-2459}
          \and
          E.~Mooij\inst{1}
          \and
          A.~Ricarte\inst{3}\fnmsep\inst{4}\orcidlink{0000-0001-5287-0452}
          \and
          D.~Palumbo\inst{3}\fnmsep\inst{4}\orcidlink{0000-0002-7179-3816}
          }

   \institute{Faculty of Aerospace Engineering, Delft University of Technology, Kluyverweg 1,                 2629 HS Delft, The Netherlands
         \and 
              Joint Institute for VLBI ERIC (JIVE), Oude Hoogeveensedijk 4, 7991 PD Dwingeloo, The Netherlands
         \and 
              Black Hole Initiative at Harvard University, 20 Garden Street, Cambridge, MA 02138, USA
         \and 
              Center for Astrophysics | Harvard \& Smithsonian, 60 Garden Street, Cambridge, MA 02138, USA
             }


 
    \abstract{ Direct observation of the orbital motion of a sub-parsec supermassive black hole binary (SMBHB) would provide the first conclusive electromagnetic proof of such systems existing. Widely considered to be the source of gravitational waves, binaries are expected to form as a natural consequence of galactic mergers, and determining the processes that drive their evolution is essential to understanding cosmological evolution. In this work, we evaluate the prospects of using ground and spaceborne Very Long Baseline Interferometry (VLBI) to observe SMBHBs and estimate their orbital parameters. The Black Hole Explorer (BHEX) is considered the primary case study. Achieving unprecedented resolution, BHEX will provide access to a new volume of binary parameter space, potentially enabling the first confident detection of an SMBHB. We present an orbit-fitting approach that uses relative astrometry and Bayesian dynamic nested sampling, and demonstrate its efficacy on a set of example binary systems. For simulating observations, we use a binary image model based on post-Newtonian orbit propagation and find that for BHEX, binary detection requires a total flux density of 40~mJy and a minimum separation of \(\sim\)2~\(\mu\)as. With three annual observations, BHEX could constrain the semi-major axis and the eccentricity of binaries with orbital periods of $\leq$10~years to within 0.06 dex of the true values under specific noise assumptions. We have also evaluated the benefits provided by BHEX for binary detection compared to ground-only observations by arrays such as the next generation Event Horizon Telescope (ngEHT). Finally, we constrained the requirements of a future spaceborne VLBI system capable of performing a statistically significant survey of SMBHBs.}
  
   {}
   {}
   {}
   {}
   {}


    \titlerunning{Towards direct imaging of supermassive black hole binaries with spaceborne VLBI}
    \authorrunning{B. Hudson et al.}

    \maketitle

%
\nolinenumbers
\section{Introduction}
\label{s:intro}

\noindent Galactic mergers are inevitable in the course of the evolution of the Universe, and supermassive black hole binaries (SMBHBs) are expected to form as a natural consequence of such mergers \citep[][and references therein]{kormendy_coevolution_2013}. Determining the processes involved in their formation and evolution is essential to 
understanding cosmological evolution on a galactic scale.

The stages of an SMBHB's life cycle (formation, inspiral, merger, and ringdown) were defined in the work by \cite{begelman_massive_1980}. Following a galaxy merger, dynamical friction is capable of driving the binaries closer together \citep{callegari_growing_2011, mayer_massive_2013, dosopoulou_dynamical_2017}. However, at a separation on the order of several parsecs, dynamical friction becomes inefficient at continuing to decay the binary orbit, a process known as hardening \citep{merritt_massive_2005}. Further evolution of an SMBHB may be driven by processes such as torque imparted on the binary due to interaction with a circumbinary gas disk (see \cite{tang_orbital_2017} and references therein) and/or asymmetric stellar distributions that promote a high interaction rate between the binary and surrounding stars \citep{gualandris_collisionless_2017}. The final decay is controlled by dissipation of the binary's kinetic energy via gravitational wave (GW) emission \citep{begelman_massive_1980, merritt_massive_2005}. Through these stages, the separation between the SMBHB components decreases from single-digit parsecs to $\lesssim0.01$~pc. Characteristic angular separations of components at these stages of evolution are of the order of 1~$\mu$as and smaller \citep{gurvits_milliarcsecond_2025}. 

From detection of the GW background, sub-parsec binaries are expected to exist, as the measurements favour models with efficient merging instead of indicating the existence of a significant final-pc problem \citep{agazie_nanograv_2023}. However, at the time of writing, there has been no conclusive direct evidence of such systems in electromagnetic observations. As described by \cite{noauthor_black_2024}, directly resolving a binary of this type is only possible through the use of Very Long Baseline Interferometry (VLBI). The angular resolution required for such a detection (and subsequent imaging) is, in the best case, near the fundamental limit of ground VLBI systems. \cite{dorazio_repeated_2018} predict that with the capabilities of the next generation Event Horizon Telescope (ngEHT), a few SMBHB systems may be resolvable out to $z\lesssim 0.2$. 

Ground VLBI systems are fundamentally limited in angular resolution by the Earth's diameter (\(\sim\)$12,000$~km) and the highest radio frequency (\(\sim\)$345$~GHz) permitted by atmospheric opacity. Thus, a robust and statistically significant investigation of multiple SMBHB objects necessitates the use of space VLBI systems, as they are free of these limitations.

Characterisation of a population of binaries would enable evolutionary models to be constrained across the SMBHB life cycle. Multi-messenger studies combining VLBI with electromagnetic observations at other wavelengths and GW observations would enormously broaden the view on the evolution of galaxy constituents of the Universe. Resolving a population of SMBHBs would also improve predictions for the stochastic GW background, breaking model degeneracies \citep{agazie_nanograv_2023}. 

Here, we consider the near-future prospects for direct resolving observations of SMBHBs with spaceborne VLBI. A motivation for this work is the prospect of synergistic studies of SMBHBs in both their electromagnetic and GW emission, the latter by pulsar timing arrays \citep{agarwal_nanograv_2026} and other future GW facilities. 

In this paper, we propose a methodology for the two objectives discussed above, namely, VLBI detection of an SMBHB and monitoring of its orbital motion using a Bayesian inference, post-Newtonian (PN) orbit fitting technique. In doing so, we provide an approach that could be used for future SMBHB VLBI observations. This technique is also used to determine the likelihood of a mission such as BHEX being able to accurately constrain orbital parameters and for defining preliminary requirements for future spaceborne VLBI systems.

In Sect. \ref{s:arrays}, we present the three future VLBI arrays used as case studies in this work. In Sect. \ref{s:conditions}, the conditions required for the proposed detection and orbit fitting approach to be applicable are discussed. The orbit fitting methodology is presented in Sect. \ref{s:method}. In Sect. \ref{s:bhex_binary}, an evaluation of the binary observation prospects of BHEX is presented, and this is accompanied by a demonstration of the orbit fitting. Finally, in Sect. \ref{s:discuss}, preliminary requirements of a future spaceborne VLBI mission that can perform imaging SMBHB surveys are defined.

\section{Future VLBI arrays}
\label{s:arrays}

\noindent The ngEHT will build on the performance of the EHT by adding more ground sites and regularly observing at 345~GHz \citep{ayzenberg_fundamental_2025}. The angular resolution of this ground array is assumed to be $15\,\,\mu$as in the subsequent analyses \citep{doeleman_reference_2023}. In addition to the ngEHT, we consider the binary-detection prospects of two proposed spaceborne VLBI instruments.

The Black Hole Explorer (BHEX) is a proposed 2-year extension to ground-based arrays such as the EHT, with an angular resolution of \(\sim\)6~\(\mu\)as \citep{johnson_black_2024}. The primary aim of BHEX is to detect the photon ring in M87* and Sgr\,A*, the horizon-scale targets of the EHT \citep{the_event_horizon_telescope_collaboration_first_2019,event_horizon_telescope_collaboration_first_2022}. On baselines with ALMA, BHEX will achieve a sensitivity of \(\sim\)1~mJy \citep{johnson_black_2024}. BHEX is intended for submission to NASA's Small Explorers (SMEX) programme, at the next call for proposals.

TeraHertz Exploration and Zooming-In for Astrophysics (THEZA) was originally prepared as a concept in response to the European Space Agency's (ESA) call for its science program Voyage~2050 \citep{gurvits_theza_2021,gurvits_science_2022}. THEZA represents a theoretical, multi-element space system, capable of using space-space VLBI to achieve an order of magnitude improvement in angular resolution. For estimation of relative astrometry errors, we assume an angular resolution of 1~\(\mu\)as for THEZA.

\section{Binary detectability conditions}
\label{s:conditions}

\begin{figure}
\centering
\includegraphics[width=\columnwidth]{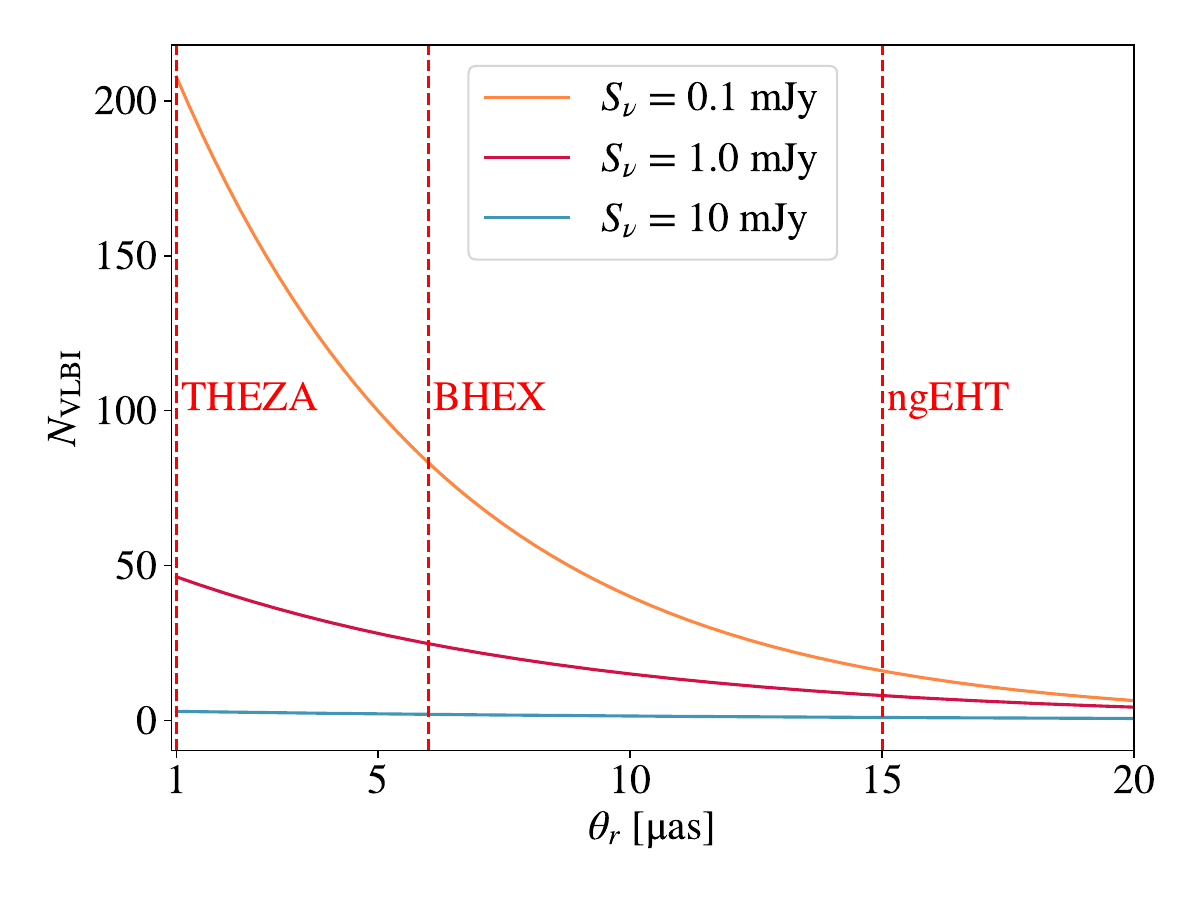}
\caption{Predicted number of observable SMBHB systems as a function of array angular resolution ($\theta_{r}$) and flux density sensitivity ($S_{\nu}$) out to $z$ = 2. Here, $P_{\text{obs}} \leq$ 10 years and $q \geq $ 0.01. The figure was generated with data from \cite{dorazio_repeated_2018}.}
     \label{f:pop}
\end{figure}

\noindent \cite{dorazio_repeated_2018} predict the number of sub-parsec binary sources observable with VLBI under the following conditions: 1. The binary orbital separation is larger than the minimum spatial resolution of the array. 2. Both components are bright enough to be independently detectable. 3. The observed orbital period is $P_\text{obs}<10$~years, which serves as a simple limit to increase the chance of detecting orbital motion. Fig. \ref{f:pop} depicts the predicted number of observable sources from their approach, with the angular resolutions of the three reference arrays plotted.

This model should be revisited given the recent GW background measurements by the NANOGrav collaboration. As a preliminary evaluation, \cite{agazie_nanograv_2023} find a strain amplitude of $2.4^{+0.7}_{-0.6} \times 10^{-15}$ for a frequency of 1 year$^{-1}$. In Fig. 2 of \cite{dorazio_repeated_2018}, the predicted GW background from their model is presented. For the same frequency, the strain amplitude broadly agrees with the pulsar timing array measurements.

Critically, the population estimate assumes that, if an active galactic nucleus has an SMBHB, both components are millimetre-bright. The results presented in Fig. \ref{f:pop} scale linearly with the value of this fraction. For example, if only 10\% have two millimetre-bright components, the predicted number of observable systems for BHEX decreases to $<10$. However, this is only for systems with $P_\text{obs}<10$~years. \cite{dorazio_repeated_2018} note that increasing $P_\text{obs}$ increases the value of $N_\text{VLBI}$, by an order of magnitude in the example they provide for $P_\text{obs}=20$~years.

Requirement 2 from \cite{dorazio_repeated_2018} is the primary assumption we adopt. \cite{ayzenberg_fundamental_2025} describe this as a telescopic or visual binary where both components have independently detectable emission. Orbit fitting can then be performed using relative astrometry, for example, as performed by \cite{fomalont_sub-milliarcsecond_1999} (i.e. estimation of the position of the secondary black hole with respect to the primary, without reference to the known position of a calibrator). The orbit fitting methodology has been developed for this specific case. However, with absolute astrometric uncertainties approaching 1~$\mu$as \citep{broderick_localizing_2011,zhao_how_2024}, this approach could in theory be used for candidate binaries with only one detectable component if its motion with respect to a calibrator source was sufficient. We do not explore this potential use case in the subsequent analyses.

Both binary components being detectable with VLBI requires conditions supporting compact radio emission, such as horizon-scale emission, such as that detected for M87* and Sgr\,A* by the EHT, and/or jet launching \citep{gutierrez_non-thermal_2024}. \cite{avara_accretion_2024} provide a summary of the conditions required for the formation of a circumbinary disk (CBD), and subsequent black hole mini-disks, from recent general relativistic magnetohydrodynamic (GRMHD) results. A CBD could form around a sub-parsec binary with a gas supply in the nucleus, where the gas has sufficient angular momentum, as is expected to occur in merged galaxies \citep{chapon_hydrodynamics_2013}. Streams can break off from the CBD's inner edge and feed accreting matter onto the binary components, with factors such as the mass ratio $q=\frac{m2}{m1}$ having a significant effect on the dynamics \citep{combi_minidisk_2022,avara_accretion_2024}. If the specific angular momentum of these streams is greater than that of the innermost stable circular orbit of the black hole, then mini-disks may form around one or both components \citep{gutierrez_non-thermal_2024}.

\cite{tiede_hot_2025} present a spectral energy distribution model for a binary system consisting of two mini-disks and a CBD. They demonstrate that their model (including a jet feature) can self-consistently capture the observed broadband spectrum of the binary candidate PG1302-102. \cite{gutierrez_non-thermal_2024} discuss the possibility of dual jet launching from sub-parsec binaries. Using RadioAstron 22~GHz observations of the famed binary candidate OJ287, \cite{valtonen_identifying_2025} show that the existence of a secondary jet from the smaller black hole is consistent with the spectral shape of the periodic optical flares observed.

Significant work remains to understand the accretion properties and thus the detectability of sub-parsec binary systems. The results of such studies have the potential to severely impact the number of observable binary systems with VLBI. However, binary observation has been identified as a potential activity of the ngEHT \citep{ayzenberg_fundamental_2025}. Within the Event Horizon and Environs (ETHER) database, used for identification of EHT and ngEHT targets, a library of SMBHB candidates for follow-up VLBI observations is also being built  \citep{ramakrishnan_event_2023}. The widespread interest in this activity warrants investigation of BHEX's binary detection prospects.

\section{Method}
\label{s:method}

\noindent The analysis pipeline presented requires a sample of relative right ascension and declination positions of the secondary black hole with respect to the primary. From the relative position data, the process is then divided into two phases: binary confirmation and orbit estimation. The former involves building confidence that what is being observed is two SMBHs exhibiting non-linear relative motion. For the latter, sufficient observations of the system over time have been performed to constrain its orbital parameters.

\subsection{Binary toy model}
\label{ss:orbit_model}

For simulating VLBI observations of SMBHBs and for calculation of the likelihood of candidate orbits in the Bayesian fitting process, we implemented a PN orbital model. The model is based on the one described by \cite{Blanchet-2024LRR} up to order 3.5PN. This is consistent with the typical consideration for massive black hole inspirals \cite[e.g.][and references therein]{piarulli_parametrized_2025}. The rational behind excluding higher-order effects is discussed below. The acceleration function is defined as

\begin{equation}
    \ddot{\textbf{x}} = -\frac{GM}{r^{2}}[(1 + \mathcal{A})\,\textbf{n} + \mathcal{B} \textbf{v}],
\end{equation}

\noindent where $M = m_{1} + m_{2}$, $\textbf{n} = \textbf{x} / r$ and the orbital separation is $r = |\textbf{x}|$. The coefficients $\mathcal{A}$ and $\mathcal{B}$, which describe the PN perturbations in increasing powers of $c$, are provided by \cite{pati_post-newtonian_2002}. Acceleration was calculated in the centre-of-mass frame, and the black hole positions are

\begin{equation}
    \mathbf{x_{1}} = \frac{m_2 \mathbf{x}}{M} \text{ and } \mathbf{x_{2}} = \frac{-m_1 \mathbf{x}}{M}.
    \label{E:bary}
\end{equation}

\noindent The acceleration function is integrated from a set of initial Campbell orbital elements defined as: semi-major axis ($a$), eccentricity ($e$), inclination ($i$), position angle of nodes ($\Omega$), argument of periastron ($\omega$) and time of periastron passage ($T_{0}$). An explicit, eighth order Runge-Kutta method was used for the integration. As described by \cite{blunt_orbits_2017}, for orbit fitting, the parameter $\tau$ is used instead of $T_{0}$, as the prior bounds for it are simply between 0 and 1, regardless of orbital period. We note that $T_0$ is related to $\tau$ by

\begin{equation}
    \tau = \frac{T_0 - t_{\text{ref}}}{P_{\text{rest}}},
\end{equation}

\noindent where $t_{\text{ref}}$ is some specified reference date and $P_{\text{rest}}$ is the rest-frame orbital period. Both are in the same time units (e.g. Modified Julian Date).

Other, higher-order effects have not been implemented for this preliminary formulation of the pipeline. Provided in Appendix \ref{A:orbit_orders} is an evaluation of the impact of these additional perturbations and justification for their inclusion/exclusion. In summary, Schwarzschild black holes are considered with zero spin. Thus, the spin-spin, spin-orbit coupling effects described in \cite{dey_authenticating_2018} are not included. Nor are the other terms in $\mathcal{A}$ and $\mathcal{B}$ beyond 3.5 PN.

The effect of the relative light travel time from each source is taken into account. Although they will be observed on a flat sky plane, any orbit that is not face-on will have one black hole further from the observer. The time taken for this extra distance to be traversed by the emission is non-negligible and will result in the farther black hole's position being from an earlier time relative to the other source. Details of how this has been modelled are provided in Appendix \ref{A:orbit_orders}.

\begin{figure}
\centering
\includegraphics[width=0.9\columnwidth]{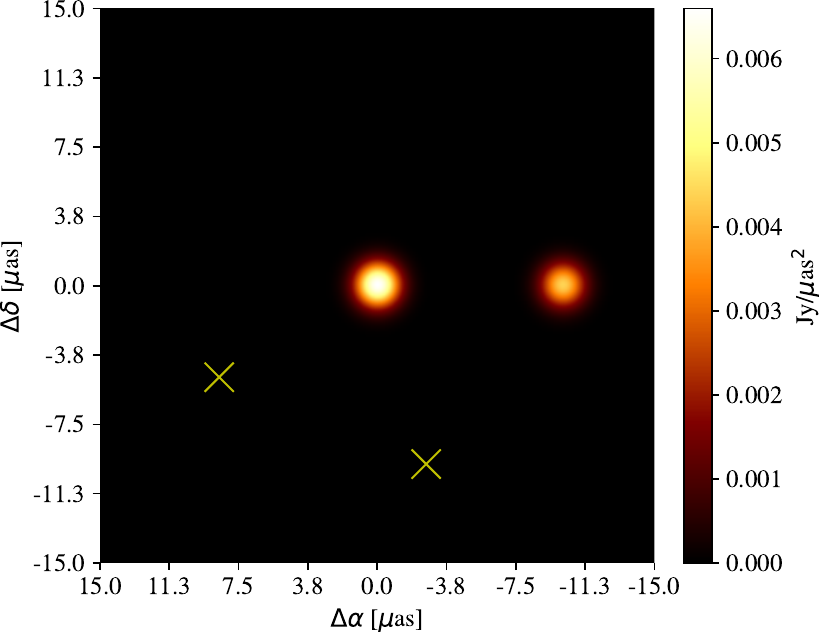}
\caption{Stokes I image of the binary toy model for Case 1 in March 2032, 2033, and 2034. The primary source is fixed to the origin and the first position of the secondary is illustrated. Subsequent observed positions of secondary are depicted with yellow crosses.}
     \label{f:case1_binary}
\end{figure}

For synthetic data simulations of BHEX observations, we build a binary image model from the superposition of two Gaussian sources, described by their Full Width at Half Maximum (FWHM) and total flux density. We note that this is a simple model of two compact radio cores, with no contributions from jets launching from one or both black holes. Use of this model also assumes that both sources are unresolved or only marginally resolved; a fair assumption as BHEX is only expected to be able to resolve an additional 5-10 black hole shadows \citep{johnson_black_2024}. We assume a flat spectrum for the components for simplicity, and because detection of such would be more difficult than for a rising spectrum. This toy model is shown in Fig. \ref{f:case1_binary}, depicting the first test case analysed in Sect. \ref{s:bhex_binary}.

\subsection{Binary confirmation}
\label{ss:orbit_conf}
Upon observing a binary candidate with a VLBI array, the challenge will be in building confidence that what has been measured is indeed an interferometric response to an SMBHB structure. With the likely faintness and small angular separation of the two sources, it is conceivable that the points of emission may not be gravitationally bound and what is being observed is two independent sources with close celestial positions or a combination of an AGN core and a jet. Other scenarios for potential false alarms can also be imagined. Confirming a binary detection will require evidence to reject these null hypotheses. This evidence can be considered in two categories:

\begin{enumerate}
    \item Confirming the sources are SMBHs: \newline For most sources of future VLBI observations of binary candidates, they will appear as unresolved and point-like or Gaussian due to the very high angular resolution required to resolve them on an event horizon (EHT-like) scale. Observational features that will provide strong evidence of the sources being black holes include: the launching of relativistic jets; spectra showing flux density rising with frequency suggesting compact, non-thermal emission; location within the centre of a known active galactic nucleus and evidence from multi-wavelength observations.

    \item Detection of orbital motion: \newline
    Confirming an SMBHB detection requires observing orbital motion. It is the process of observing and constraining this motion that the rest of this section is focused on.
\end{enumerate}

\noindent We consider linear motion as the null hypothesis to be rejected in the attempt to detect orbital motion. This is based upon the assumption that if the objects are not gravitationally bound, they are most likely to be exhibiting linear motion on the small angular scales a VLBI mission can probe. Systems on much larger orbital paths, which could be the source of erroneous binary detections, would likely appear as linear motion on these scales. Other sources of erroneous binary detection can be imagined, for example, bright components in a relativistic jet exhibiting some relative motion on short timescales. In such cases, where it is feasible that relative motion may be non-linear, the other forms of evidence listed above would have to be carefully considered along with the goodness of fit of an orbit, determined using this approach.

A useful metric for planning future VLBI missions is the number of observations required to confidently detect curvature in the observed motion. This expression is based on the need for the curvature in the trajectory to exceed the uncertainty in the position measurements. For an inclined circular orbit, the number of observations, $N_{min}$, required is

\begin{equation}
    N_{min} = \max\left( 3,\ \Biggl \lceil 1 + \sqrt{ \frac{k \sigma_{\text{pos}} P_{\text{obs}}^2}{a \pi^2 t_{\text{cad}}^2 \cos{i}} }  \Biggr \rceil \right),
\label{E:min_n}
\end{equation}

\noindent where $P_{\text{obs}}=(1+z)\,P_{\text{rest}}$, $t_{\text{cad}}$ is the cadence of observations, and $a$ is the orbit semi-major axis in angular units. Further, $k$ is a significance threshold, which we set to 5, and $\sigma_{\text{pos}}$ is the uncertainty on the position estimation. Appendix \ref{A:linear_motion} contains the full derivation of this analytic expression. 

To perform relative astrometry where two sources are within the primary beam (as will certainly be the case for sub-parsec binaries), \cite{thompson_interferometry_2017} provide the following equation in Appendix 12.1.3 for the thermal noise limited position uncertainty,

\begin{equation}
\sigma_\text{pos} \simeq \frac{\sigma_{\phi}}{2\pi \sqrt{N} D_{\lambda}}.
\end{equation}

\noindent Approximating the phase error $\sigma_{\phi}$ as $1 / $ signal-noise ratio (S/N) for a strong detection, with $\text{S/N}=5$, and converting to angular resolution $\theta_\text{r} = 1 / D_{\lambda}$,

\begin{equation}
\sigma_\text{pos} \simeq \frac{\theta_\text{r}}{2\pi \sqrt{N} \, \text{S/N}},
\label{e:sigma_pos}
\end{equation}

\noindent where $N = \frac{N_A(N_A - 1)}{2}$ is the number of independent baseline measurements, and $N_A$ is the number of antenna in the array.

\begin{figure}
\centering
\includegraphics[width=\columnwidth]{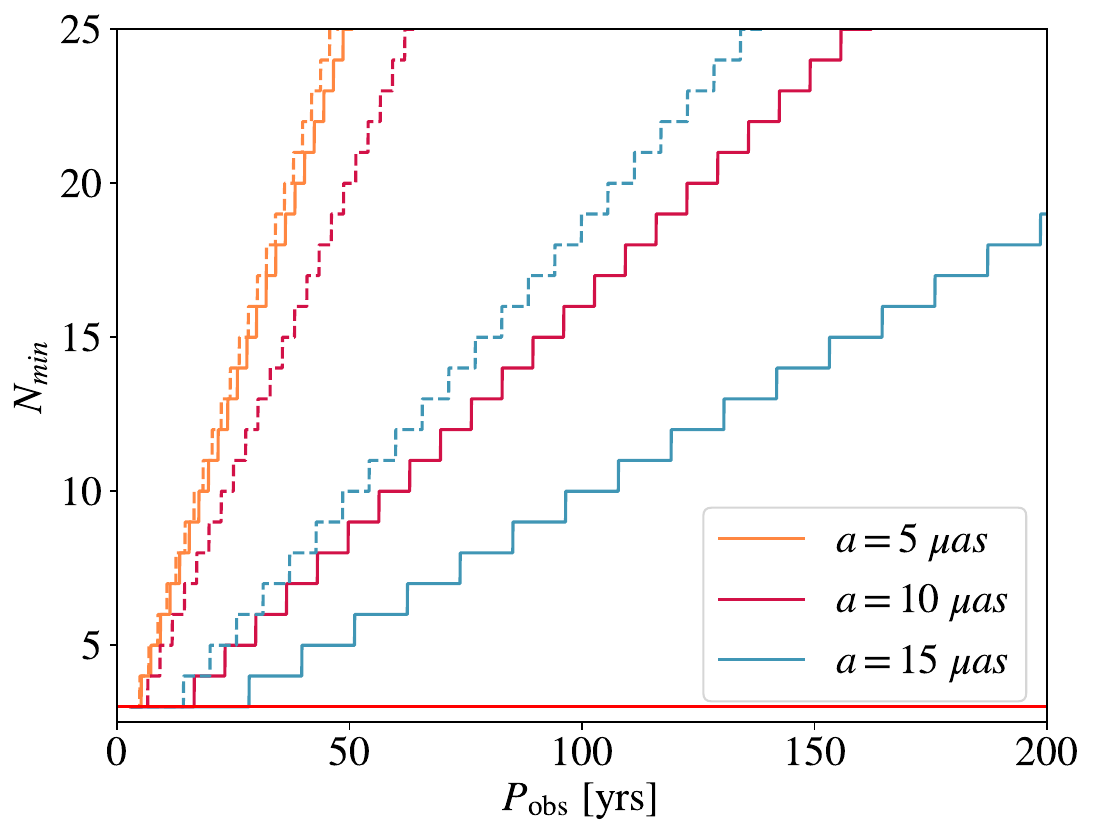}
  \caption{Minimum number of observations required to detect the curved motion of the secondary black hole as a function of the observed orbital period and semi-major axis in angular units. Observations are performed annually. We assume the nominal angular resolution of 6~\(\mu\)as for BHEX (solid lines), and 15~\(\mu\)as for the ngEHT (dashed lines). The red line depicts the assumption used in synthetic data simulations: the best case for the number of BHEX observations (three) of a binary across a nominal 2~year mission.}
     \label{f:Min_N}
\end{figure}

Fig. \ref{f:Min_N} depicts $N_{\text{min}}$ for a circular, face-on orbit of varying semi-major axis and orbital period (Fig. \ref{f:pop} can be used to relate these angular separations to the population estimates of \cite{dorazio_repeated_2018}).

As discussed in Sect. \ref{ss:synth_data}, Eq. \ref{e:sigma_pos} does not include the effect of additional positional uncertainty introduced by model misspecification and systematic error. Fig. \ref{f:Min_N} has been produced using the values of $\sigma_{\text{pos}}$ from Table \ref{t:error_survey}, describing the position uncertainties achieved in synthetic data simulations of BHEX and the ngEHT, for the 10\% + 10~mJy error case. For reference, the thermal noise limited expression gives estimates of the relative astrometric uncertainty of  $0.005~\mu$as, $0.032~\mu$as and $0.080~\mu$as at $N_A=9$, for THEZA, BHEX and ngEHT, respectively.

Fig. \ref{f:Min_N} shows that for three annual BHEX observations of a system across a 2~year mission, curved trajectory motion could be confidently detected in binaries with $5<a\,[\mu\text{as}]<15 $ and $7<P_\text{obs}\,\text{[years]}<28$. The observational limits associated with BHEX are of course removed for binaries that are detectable with the ngEHT, and the effect of increased astrometric uncertainty without BHEX can be seen in Fig. \ref{f:Min_N}. Assuming similiar sensitivity to BHEX, with a 1~$\mu$as resolution, THEZA should provide a 6 times improvement in astrometric uncertainty, and \(\sim\)2.5 times lower $N_{min}$, to detect the same orbit.

Extending this approach to elliptical orbits as part of the Bayesian orbit fitting approach, the false alarm rate (FAR) of the maximum a posteriori (MAP) solution is calculated. This FAR represents the probability that the apparent orbit motion signature could simply arise from noise in a linear trajectory. A two-parameter linear model is fitted to the data (gradient and intercept in the ($\Delta \alpha$) and ($\Delta \delta$) system). The FAR is then calculated using a chi-squared test.

\subsection{Orbit estimation}
\label{ss:orbit_det}

Once sufficient observations of a source have been performed to gain confidence that it is a likely binary candidate, the orbital parameters can be estimated. \cite{fang_orbit_2022} present a Newtonian orbit fitting approach for a binary model consisting of two point sources. They demonstrate how the orbital parameters can be determined directly from the visibility data. Here we present an extended method, consisting of a PN orbit propagation and one that is applicable to various forms of binary models (i.e. not just point sources) and can also be used on reconstructed images.

The group of Bayesian Markov chain Monte Carlo (MCMC) methods is widely used in astronomy for orbital parameter estimation across a diverse range of disciplines, from stellar binaries to exoplanets \citep{pearce_orbital_2020,blunt_orbits_2017}. Other algorithms such as the least squares Monte Carlo method or more traditional orbit fitting (e.g. Gauss's and Gibbs methods, Lambert's problem) are effective at finding only the best-fit solution under specific required conditions for the number of and cadence of observations.

However, as noted by \cite{thompson_octofitter_2023}, it is challenging to compute posteriors for traditional orbital elements as they consist of complex co-dependencies and degeneracies. This is exacerbated when working with short orbital arcs, as is likely to be the case for SMBHB observations. Depending on the type of MCMC method used, they can be ineffective at estimating the posterior for highly multi-modal problems. Variants such as the parallel-tempered method used by \cite{blunt_orbits_2017} aim to address this issue.

Dynamic nested sampling has been shown to be effective at estimating Bayesian posteriors in an astronomical context, particularly when compared to MCMC methods \citep{speagle_dynesty_2020}. Dynamic nested sampling adaptively allocates samples depending on the posterior structure, enabling it to effectively sample multi-modal distributions. For this orbit fitting approach, we have implemented the dynamic nested sampling Python package \texttt{dynesty}\footnote{\url{https://github.com/joshspeagle/dynesty}} \citep{speagle_dynesty_2020}. A custom, Gaussian log-likelihood function was used of the form

\begin{equation}
    \log{\mathcal{L}(\mathcal{D} \mid m)} = -\frac{1}{2} \sum_{i=1}^{N_{\text{obs}}}{\frac{(\Delta \alpha_{obs,n} - \Delta \alpha_{m,n})^2}{\sigma_{\text{pos},n}^2} + \frac{(\Delta \delta_{obs,n} - \Delta \delta_{m,n})^2}{\sigma_{\text{pos},n}^2}},
\end{equation}

\noindent where the subscript $obs,n$ are the observed positions of the $n^\text{th}$ observation and $m,n$ are the positions calculated from the trial model parameters using the PN orbit model. $N_{\text{obs}}$ is the number of observations. A prior transform is used that only allows physically realistic values of the orbital parameters and black hole masses. This is discussed in Sect. \ref{ss:param_const}.

The orbit fitting calculates the optimum values of the following parameters to fit the observational data: total mass ($M$), mass ratio ($q$), $a$, $e$, $i$, $\Omega$, $\omega$, and $\tau$. The tuneable parameters of \texttt{dynesty} have been determined through preliminary runs of the test cases presented below. We use the random walk sampling method and multiple ellipsoid bounding, as is suggested for multi-modal distributions. For live points, \cite{speagle_dynesty_2020} suggests 50 times the number of degrees of freedom, for each expected mode. Through preliminary tests, increasing the value of \texttt{nlive\_init} beyond 12000 produced minimal change in already smooth and well-sampled posterior contours, and therefore this value was used for all cases. The \texttt{dlogz} tolerance, the stopping condition for the initial baseline run, is set to 0.005. 

\subsection{Parameter constraints}
\label{ss:param_const}

\noindent An upper limit can be applied to the black hole mass ranges based on whether the two binary components have been resolved in the VLBI observations. The diameter of a resolved black hole shadow is related to its mass ($m$) by

\begin{equation}
    \theta_s \approx \frac{2\sqrt{27} Gm}{c^2D_A},
\end{equation}

\noindent where $D_A$ is the angular diameter distance, which can be calculated from knowledge of the source's redshift \citep{Bardeen:1973tla}. Redshift can be measured from other observations of the host galaxy's emission or absorption lines. As such, even if the shadow is not resolved, this expression provides an upper limit on the black hole mass. A lower limit can be estimated from spectral energy distribution models of such systems \citep{pesce_toward_2021,tiede_hot_2025} and the fact that the source has been detected in the first place. Scaling relations between the black hole and host galaxy will also be useful for determining the likely lower limit. For the subsequent example cases, it is assumed that the total black hole mass is known to within an order of magnitude of solar masses.

Semi-major axis can be constrained from measurement of the angular separation of the binary across different observations. Assuming a near face-on orbit, the maximum semi-major axis can be determined by assuming a highly elliptical orbit and that the observed black holes are currently at periastron (where $r_\text{sep}$ is the observed separation at this point).

\begin{equation}
    a_{\text{max}} = \frac{r_{\text{sep}}}{1-e}
    \label{E:max_a}.
\end{equation}

\noindent For an inclined orbit, the observed separation is a function of the semi-major axis and the inclination. In the worst-case, a binary may be viewed completely edge on ($i = \frac{\pi}{2}$) with the black holes almost aligned. The observed instantaneous separation would then be a tiny fraction of the true semi-major axis. From multiple observations (as would be required to perform orbit determination), constraints will be able to be applied to the semi-major axis. The prior imposed on $a$ ranges from 0.001~pc to twice the true semi-major axis.

Eccentricity is allowed to vary between 0 and 1, permitting only elliptical orbits to be included in the orbit fitting. Although hyperbolic orbits where the black holes are not yet fully bound are theoretically possible, they are far less likely on the small angular scales being probed with VLBI. It is also less probable to observe such a system, given how little time it would spend in such a state, when considered on cosmological timescales. The angular orbital elements are allowed to vary within their full range in the subsequent examples. The epoch of periastron passage ($\tau$) is allowed to vary between 0 and 1, as described in the previous section. 

\subsection{Synthetic data simulations}
\label{ss:synth_data}

\noindent To demonstrate the orbit fitting approach and evaluate BHEX's efficacy at detecting binaries, simulated BHEX observations have been modelled using \texttt{eht-imaging} \footnote{\url{https://github.com/achael/eht-imaging}} \citep{chael_interferometric_2018} and \texttt{ngehtsim} \footnote{\url{https://github.com/Smithsonian/ngehtsim}} \citep{pesce_atmospheric_2024}. These packages propagate the spacecraft and ground array positions across the simulation time and simulate VLBI observations of a given image model. Impacting factors such as thermal noise, weather conditions, atmospheric effects and station-based gain variations are included. \texttt{ngehtsim} provides the ability to model frequency phase transfer (FPT), allowing the phase from a lower frequency observation to be transferred to the higher band to increase coherence time, which is intended to be performed for BHEX and ngEHT \citep{rioja_exploration_2011, pesce_atmospheric_2024,issaoun_first_2025}. See Appendix \ref{As:sim_config} for a full definition of the ground array and other simulation configuration parameters.

The toy binary model presented in Sect. \ref{ss:orbit_model} is used as the image model for each of the test cases examined below. Once simulated observations have been performed, we use the native model fitting functionality of \texttt{eht-imaging} to fit two Gaussian sources to the simulated visibility amplitudes and closure phases. We fix the brighter Gaussian component to the origin such that it becomes the astrometric phase origin. A flat prior is defined for the secondary component's position between $\pm$50~\(\mu\)as. A flat prior is used for both component's total flux density and FWHM of 0--1~Jy and 0--30~\(\mu\)as, respectively. From the model fitting, we extract an estimate of $\sigma_\text{pos}$, the uncertainty in the relative position estimate of the secondary with respect to the primary. We also fit a single Gaussian source to the same data and evaluate the difference in the goodness of fit to assess the likelihood of a false detection of a binary.

As the binary toy model consists of idealised, compact Gaussians, it does not include the effect of model misspecification, where extended or evolving source structure would introduce additional structural phase contributions and thus increase the position uncertainty. We preliminarily constrain the effect of this by introducing varying fractional and flat additive noise, for different binary separations, and evaluate the relative astrometric uncertainty using the synthetic data simulations. Table \ref{t:error_survey} presents these results.

Table \ref{t:error_survey} demonstrates the expected inflation of relative astrometric uncertainty with increasing fractional and additive noise. For the low noise cases, the uncertainty approaches the thermal noise limited value calculated in Sect. \ref{ss:orbit_conf}. We associate the fractional noise with systematic errors, such as those presented in the investigation conducted for EHT observations of Sgr\,A* by \cite{event_horizon_telescope_collaboration_first_2022_II}. The additive noise accounts for model misspecification effects expected in real observations. For the subsequent test cases, we adopt a 10\% fractional noise, inline with the upper limit presented by \cite{event_horizon_telescope_collaboration_first_2022_II} for the EHT. We apply an additive 10~mJy noise, which equates to a 20\% model misspecification error, for the 50~mJy total flux sources we use in the majority of the test cases. Provided in Sect. \ref{sss:add} is a case in which 10\% + 20~mJy is assumed, to demonstrate the effect of inflating noise on the orbit fitting.

\renewcommand{\arraystretch}{1.3}
\begin{table}
\caption[]{Relative astrometric uncertainty noise survey.}
\centering
\begin{tabular}{c|c c c}
\hline
$\sigma_\text{pos}$ [$\mu$as] & \multicolumn{3}{c}{Separation [$\mu$as]} \\
\cline{1-4}
Noise Case & 5 & 10 & 15 \\
\hline
0\% + 0 mJy  & 2.19 / 0.06 & 0.04 / 0.03 & 0.02 / 0.02\\
0\% + 5 mJy  & 2.62 / 1.50 & 0.38 / 0.12 & 0.23 / 0.13\\
0\% + 10 mJy  & 2.62 / 2.27 & 2.57 / 0.39 & 2.57 / 0.23\\
\hline
10\% + 0 mJy  & 2.56 / 0.28 & 0.21 / 0.08 & 0.11 / 0.07\\
10\% + 5 mJy  & 2.58 / 1.58 & 0.59 / 0.13 & 0.26 / 0.14\\
10\% + 10 mJy  & 2.61 / 2.31 & 2.85 / 0.45 & 0.91 / 0.23\\
\hline
20\% + 0 mJy  & 2.52 / 1.00 & 0.62 / 0.11 & 0.20 / 0.12\\
20\% + 5 mJy  & 2.54 / 1.77 & 1.37 / 0.15 & 0.33 / 0.17\\
20\% + 10 mJy  & 2.67 / 2.39 & 3.62 / 0.52 & 1.40 / 0.25 
\label{t:error_survey}
\end{tabular}
\tablefoot{FWHM of both sources is fixed at 2~$\mu$as. The combined total flux density is set to 50~mJy, inline with BHEX's minimum detectability condition (see Fig. \ref{f:bhex_binaries}). In each cell, the relative astrometric uncertainty is provided for ground / ground + BHEX.}
\end{table}

The systematic noise survey demonstrates how model misspecification errors can cause non-detections of close binaries. Noting in particular the 5~$\mu$as case, even small amounts of additive noise reduce detection confidence from nearly 100$\sigma$, to a $\lesssim3\sigma$ marginal detection. Table \ref{t:error_survey} also shows the benefit of adding BHEX to the ground array. For binaries with separations on the order of ground ngEHT resolution, BHEX provides a factor of 3--4 improvement on relative astrometric uncertainty. As the binary separation is reduced (as is likely for real systems), the finer resolution of BHEX provides greater benefit. For example, leading to improvements by factors of up to 9--10, under various noise conditions at 10~$\mu$as separation.


\section{SMBHB detection with BHEX}
\label{s:bhex_binary}

\noindent BHEX will utilise a 3.4~m antenna and observe in collaboration with a ground array at an upper ($\sim$320~GHz), and lower ($\sim$86~GHz) frequency band. The spacecraft will operate in a polar, medium Earth orbit (MEO), at an altitude of \(\sim\)26562~km. Justification for this orbit selection is given by \cite{hudson_toward_2025}. The reader is referred to \cite{johnson_black_2024} and the accompanying SPIE papers for more information on the design and operation of the spacecraft. 

In this section, the binary detection prospects of BHEX are evaluated. We conduct a preliminary assessment of the detectable binary parameter space, and demonstrate the orbit fitting methodology using synthetic data simulations of BHEX observations for a number of example test cases.

\subsection{Binary detection}
\label{ss:bhex_detections}

\begin{figure*}
\centering
\includegraphics[width=\textwidth]{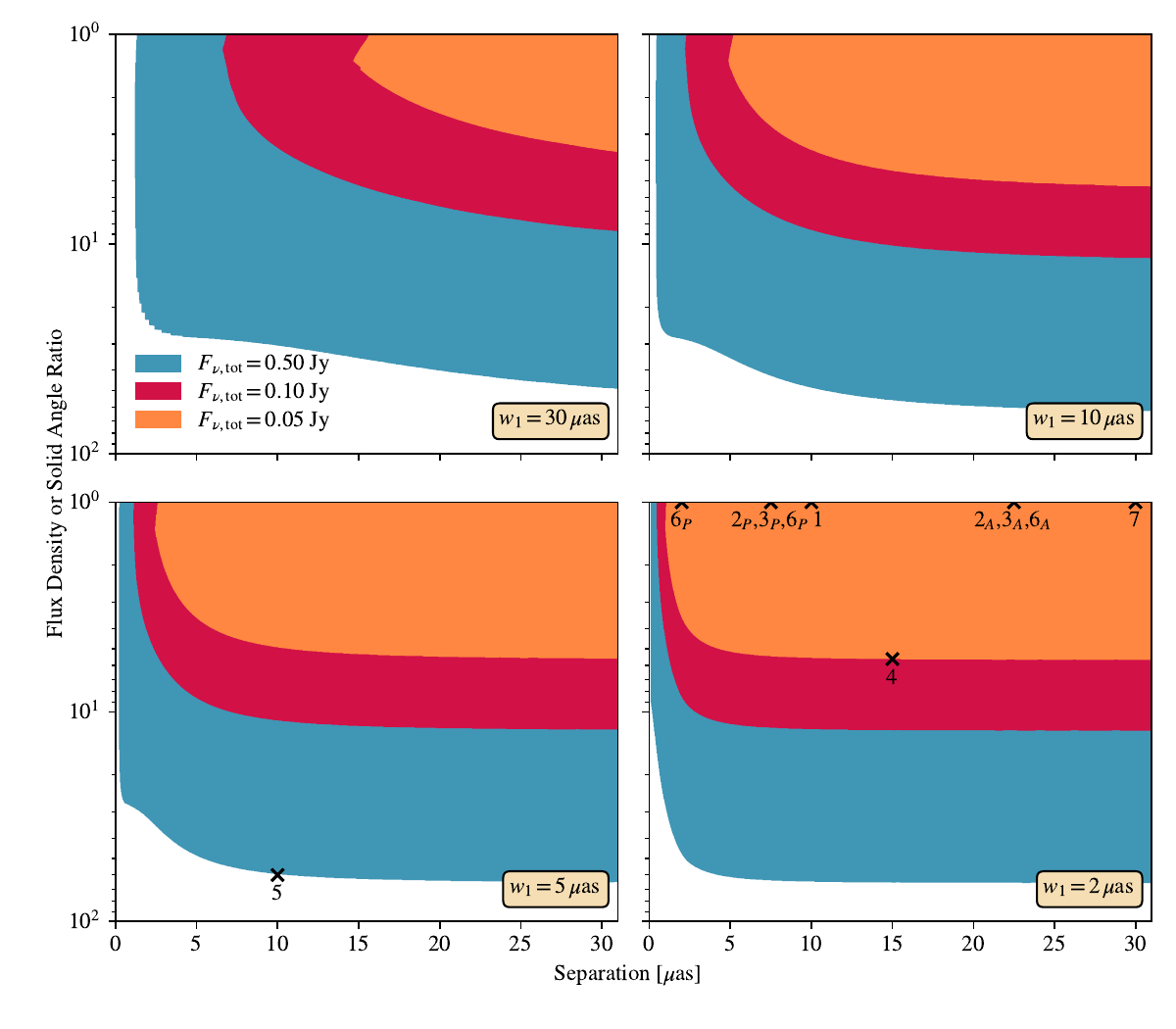}
  \caption{Binary system detection with BHEX depicting the total flux density of the source required to distinguish a binary from a single Gaussian source, assuming a thermal noise of $5$~mJy/beam. The term $w_1$ is the angular diameter of the primary component. The y-axis describes the ratio between the primary and secondary component flux densities and solid angles. For illustrative purposes, we assumed the same brightness for both components of the binary system. The positions of the seven test cases presented below are labelled in this parameter space. Subscript $A$ indicates the largest angular separation of the orbit from the observer's perspective, and $P$ is the smallest observed separation.}
     \label{f:bhex_binaries}
\end{figure*}

As an interferometer, BHEX and a ground array of telescopes samples the complex visibility of a target source -- the noise-corrupted Fourier transform of its true image. Detecting a binary first requires discerning two distinct sources in the visibility data.  Fig. \ref{f:bhex_binaries} shows the observable region of binaries by BHEX, as a function of flux density and solid angle ratio ($f$), and separation between the primary and secondary components. This figure uses the binary toy model of two Gaussians described in Sect. \ref{ss:orbit_model}. The figure depicts the required total flux density of the binary for BHEX to distinguish it from a single Gaussian source. 

A binary detection is defined as when the correlated flux density on BHEX baselines is $3\sigma$ discrepant from the total flux density and also from the flux density of the primary Gaussian source. Fig. \ref{f:bhex_binaries} also depicts the locations of the test cases analysed in Sect. \ref{ss:bhex_orbital} in the parameter space. The subscript refers to their angular separation at apoastron ($A$) and periastron ($P$). This figure only includes the effects of thermal noise, and as such should be used as a preliminary guide to BHEX binary detectability, before synthetic data simulations are performed of specific cases.

BHEX is able to detect binaries with a lower total flux density as the primary becomes more compact (moving from left to right in Fig. \ref{f:bhex_binaries}) and the separation increases. $F_{v,tot}=0.5$~Jy is the approximate total flux density of M87*, the brightest horizon-scale source for ground VLBI observations. As such, the white regions of this figure depict parameter space where a total flux density greater than this is required to detect a binary. Therefore, it becomes increasingly less likely that such systems exist. Considering this constraint, BHEX is able to detect binaries with a flux density ratio of $f \geq 0.045$, if it is assumed sources with $F_{v,tot}=0.5$~Jy can be found. 

Assuming a thermal noise figure of $5$~mJy (representative of baselines across the array), binaries with a total flux less than $40$~mJy are not detectable, noting that the thermal noise is dependent on the specific ground sites participating in the observation. In the best case on BHEX-ALMA baselines, binaries with a total flux of $\geq 7$~mJy are detectable, for $f=0.75$. As has been found in GRMHD simulations of close binary systems, for low mass ratios the secondary may be more active than the primary \citep{dorazio_accretion_2013,farris_binary_2014, dorazio_transition_2016}. It is therefore non-trivial to relate $f$ to a specific, detectable mass ratio. BHEX data can be used to identify the presence of a binary source down to separations of \(\sim\)2~\(\mu\)as, below the nominal array resolution. 

\subsection{Orbital parameter estimation}
\label{ss:bhex_orbital}

\noindent In this section, we demonstrate the orbit fitting approach presented in Sect. \ref{s:method}, and evaluate BHEX's ability to detect orbital motion in binaries. For a number of test cases, the estimated positions (and uncertainties) of the secondary with respect to the primary component were determined from the synthetic data simulation pipeline. We then ran the orbit fitting approach on these data to estimate the orbital parameters of the system.

The test cases were selected to explore as much of the binary parameter space as possible, within the limits imposed by the computational expense of running these simulations. We focused on the minimum and maximum of important parameters, such as total flux density, inclination, angular separation, and orbit period. Three primary test cases were selected:

\begin{enumerate}
    \item A circular face-on binary with a short $P_{\text{obs}}$ to demonstrate a system for which a large percentage of the orbital period is sampled. The FWHM of each Gaussian component is set below BHEX angular resolution, as is likely for a binary at redshift higher than M87* and Sgr\,A*;
    \item An eccentric, face-on binary to test the effect of eccentricity on the orbit fitting. The periastron separation is set near the limit of BHEX angular resolution, with a longer $P_{\text{obs}}$ (for which observable systems are more likely to exist) to demonstrate the increased uncertainty in fitting an orbit to a short observed arc;
    \item An eccentric, inclined binary to test the effect of inclination on the orbit fitting.
\end{enumerate}

\noindent For all three, $m_1$ and $m_2$ are equal but the flux density is unevenly distributed between the components to break the degeneracy that exists in the visibility domain if the sources were identical. A total flux density of 50~mJy is used, in accordance with the limit of BHEX's ability presented in Fig. \ref{f:bhex_binaries}, for the upper and lower bands' combined total flux density of the sources.

It is also ensured that the test cases stay within realistic limits for mass-luminosity relations and brightness temperature. The luminosities of the components in each test case have been evaluated, and compared to the millimetre-wavelength fundamental plane presented by \cite{ruffa_fundamental_2023}, for a fixed redshift of $z=0.05$. The brightness temperature ($T_\text{b}$) of each component has also been calculated. For reference, for the primary source used in the first three examples with $F_{\nu,\text{tot}}=0.03$~Jy, $T_\text{b}=9.4\times10^{10}$~K, within the typical brightness temperature range of VLBI sources.

In a nominal \(\sim\)2~year mission, the first 1--2 months will be spent in a commissioning phase. For the subsequent examples, an observing cadence of $t_{\text{cad}}=1$~year is assumed, beginning as soon as commissioning is complete with two subsequent observing sessions of the same source, during the M87* observation window. The source is observed for one night at each of these epochs and is assumed to lie close to M87* on the sky.

Summaries of the key properties of each test case, and the results of the synthetic data simulation and orbit fitting, are presented in Table \ref{t:test_summary}. A number of additional test cases are more briefly analysed in Sect. \ref{sss:add}, to further explore the effect of key parameter variations.

\renewcommand{\arraystretch}{1.3}
\begin{table*}
\caption[]{Summary of the test cases.}
\centering
\begin{tabular}{c c c c c c c c c c c}        
\hline
Case & $F_{\nu,\text{tot}}$ [Jy] & $q$ & $P_{\text{obs}}$ [yr] & Separation [\(\mu\)as] & $e$ [-] & $i$ [$\degree$] & $M$ [$10^8$ \(\textup{M}_\odot\)] & $\text{FWHM}_{1/2}$ [\(\mu\)as] &  \\
\hline 
1 & 0.03/0.02 & 1 & 5 &  10 & 0 & 0 & 4.00 & 2/2  \\
2 & 0.03/0.02 & 1 & 10 & $r_{p}=7.5$, $r_{a}=22.5$ & 0.5 & 0 & 3.37 & 2/2  \\
3 & 0.03/0.02 & 1 & 10 & $r_{p}=7.5$, $r_{a}=22.5$ & 0.5 & 45 & 3.37 & 2/2   \\
\hline
& $\nu L_{\nu}$ [ergs$^{-1}$] & $\chi^2_{amp}$ & $\chi^2_{cp}$ & $\chi^2_{amp,sin}$ & $\chi^2_{cp,sin}$ & $\sigma_\text{pos,av}$ [\(\mu\)as] & $\chi^2_{orb}$ & FAR [\%]  \\
\hline
1 & 6.07/4.05$\times 10^{41}$ & 0.06 & 0.04 & 0.72 & 0.10 & 0.95 & 0.01 & 11  \\
2 & 6.07/4.05$\times 10^{41}$ & 0.06 & 0.04 & 0.63 & 0.31 & 0.33 & 0.02 & 0 \\
3 & 6.07/4.05$\times 10^{41}$ &  0.05 & 0.05 & 0.48 & 0.22 & 0.72 & 0.02 & 0 \\
\hline
\label{t:test_summary}
\end{tabular}
\tablefoot{For all cases, $\Omega$, $\omega$, and $\tau$ were set to zero, as they are expected to have less impact on the orbit fitting than the varied parameters. The terms $\chi^2_{amp}$ and $\chi^2_{cp}$ describe the average goodness of fit of the binary Gaussian model in amplitude and closure phase to the observed data in \texttt{eht-imaging}, across the three observation epochs. The subscript $sin$ describes the $\chi^2$ values for fitting a single Gaussian source to the observed data, and $\chi^2_{orb}$ describes the goodness of fit of the MAP orbital solution.}
\end{table*}

\subsubsection{Case 1: Circular, minimum separation}
\label{sss:case1}

Table \ref{t:test_summary} and \ref{t:case1_summary} summarise the model and orbit fitting results. The true, median and MAP orbit parameters are provided in Table \ref{t:case1_summary}, along with relative errors where they can be calculated. Fig. \ref{f:case1_bestfit} depicts the best-fit orbit. The posterior distribution corner plot is provided in Appendix \ref{As:supp}.

As can be seen in all test cases, the small number of observations results in a highly multi-modal problem. However, important constraints can be pulled from the posteriors. Total mass ($M$) has a single main mode, and it shows a relatively tight posterior distribution. As a large arc of the orbit has been sampled, the semi-major axis ($a$) and eccentricity ($e$) are very well constrained, with a clear tendency towards a circular orbit. The mass ratio ($q$) is degenerate without prior constraints on the mass distribution, which we have not imposed. We also see the expected degeneracy between $M$ and the angular parameters $i$, $\Omega$, and $\omega$, as the former controls the dynamical scale of the orbit, and the latter controls the geometric projection on the sky. In particular, posteriors for $\Omega$ and $\omega$ are expected to be wide as for a circular orbit, they do not change its projected shape. The effect of angle wrap-around and a 180\(\degree\) degeneracy in the parameters is also evident, and this is discussed further in the next section. The posterior distribution suggests a low inclination, but in general it is poorly constrained. $\tau$ is showing some tendency towards 0 or 1, and this is expected as both describe the periastron position at which observations began. We see a broader posterior in $\tau$ compared to latter cases as for a nearly circular orbit, the periastron position is weakly defined.

An interesting phenomena can be observed at the lowest observed point, at which $\sigma_\text{pos}$ is considerable larger than for the other observations. Near $\Delta\alpha=0$, the primarily east-west distribution of ground \emph{(u,v)} coverage, and the position of BHEX at the start of observations in the simulations (also aligned east-west), provides weak constraints on the $\Delta\delta$ position of the secondary. This can also be observed in Case 3, and in general we have determined that it occurs in this plane when angular separations are $\lesssim10\,\mu$as. Despite this, the MAP solution traces the true orbit very closely as the narrower error bars on the other points strongly pin the dynamics. The increased uncertainty on this data point results in a non-negligible FAR of 11\%.

In general, the posterior shows a circular, low inclination  orbit with a well-constrained dynamical scale, but with strong degeneracies and multimodality in angular parameters.

\begin{figure}
\centering
\subfigure{\includegraphics[width=0.85\columnwidth]{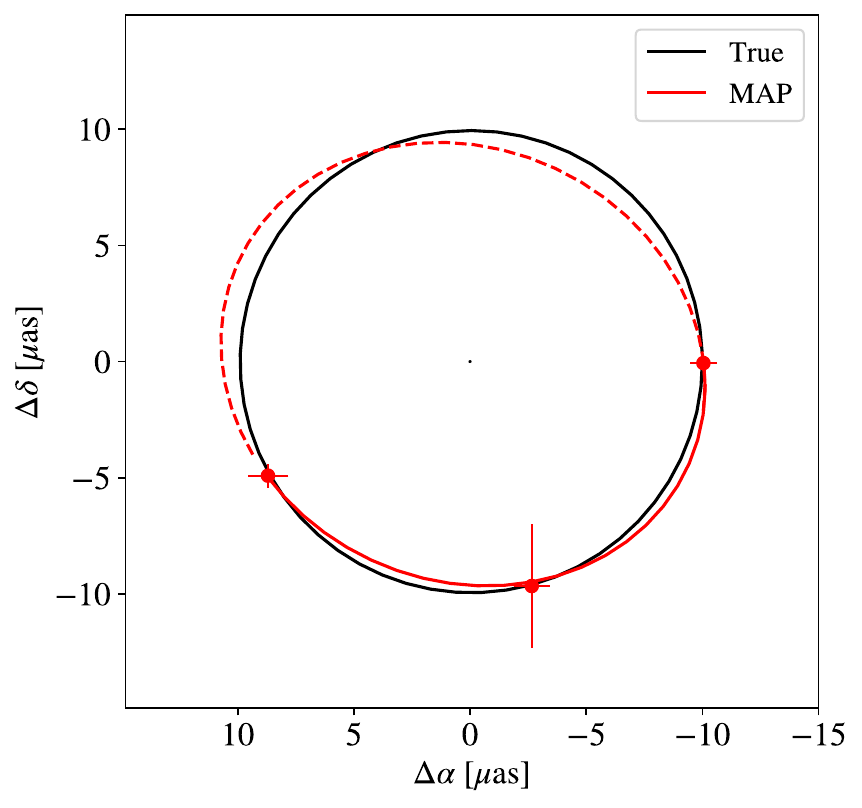}} 

  \caption{Best-fit solution for Case 1. The observed positions of the secondary relative to the primary are indicated by red markers with 1$\sigma$ error bars.}
     \label{f:case1_bestfit}
\end{figure}

\renewcommand{\arraystretch}{1.3}
\begin{table}
\caption[]{Case 1 orbit fitting summary.}
\centering
\begin{tabular}{c c c c}
\hline
Parameter & True & MAP & Median \\ 
\hline
$M$ [$10^9$~\(\textup{M}_\odot\)] 
& 0.400 
& 0.528 (32.0\%) 
& $0.543^{+0.355}_{-0.134}$ \\

$q$ [-] 
& 1.000 
& 0.924 (7.64\%) 
& $0.418^{+0.332}_{-0.283}$ \\

$a$ [pc] 
& 0.0101 
& 0.0108 (6.90\%) 
& $0.0110^{+0.00161}_{-0.00081}$ \\

$e$ [-] 
& 0.000 
& 0.0380 
& $0.085^{+0.204}_{-0.053}$ \\

$i$ [$^\circ$] 
& 0 
& 28.8 
& $28.2^{+17.3}_{-13.7}$ \\

$\omega$ [$^\circ$] 
& 0 
& 314.3 
& $214.7^{+80.7}_{-131.9}$ \\

$\Omega$ [$^\circ$] 
& 0 
& 24.2 
& $179.7^{+94.9}_{-144.4}$ \\

$\tau$ [-] 
& 0 
& 0.951 
& $0.331^{+0.493}_{-0.232}$ \\
\hline
\label{t:case1_summary}
\end{tabular}
\tablefoot{Error between fitted value and true parameter value given in brackets for MAP. Median provided with 68\% credible interval ranges.}
\end{table}

\subsubsection{Case 2: Eccentric, longer period}
\label{sss:case2}

Table \ref{t:test_summary} and \ref{t:case2_summary} summarise the model and orbit fitting results for the second test case. Fig. \ref{f:case2_bestfit} depicts the best-fit orbit. The posterior distribution corner plot is provided in Appendix \ref{As:supp}.

With the longer period orbit, a smaller arc has been sampled by the observing cadence. Naturally, this widens the posterior distribution on the orbital parameters and the relative errors of the MAP. Although across a 3~year period the trajectory exhibits less curvature than in Case 1, the narrow 1\(\sigma\) error bars still result in a confident detection of a curved trajectory, with the FAR near 0.

Semi-major axis ($a$) and eccentricity ($e$) are still very well constrained. Despite $<\frac{1}{3}P_\text{obs}$ having been sampled, a strong preference for an eccentricity \(\sim\)0.5 can be seen. Mass ratio ($q$) is unconstrained by the data without further information or assumptions, and this is evident throughout the test cases. As with Case 1, the posterior distribution of $i$ suggests a low inclination but it is very wide, with the MAP and median values \(\sim\)40$\degree$. In $\Omega$ and $\omega$ (the true values of which are both 0\(\degree\)), multiple effects can be seen: 180\(\degree\) degeneracy such that $(\omega, \Omega) \rightarrow (\omega + 180\degree, \Omega + 180\degree)$, demonstrated by the MAP solution where both parameters are \(\sim\)180\(\degree\) offset from the truth, and  wrap-around as expected for parameters ranging from 0--360\(\degree\). The epoch of periastron passage ($\tau$) is showing a strong preference for 0 and 1, both of which describe the same position. 

The posterior indicates a strong eccentric mode with low inclination and a well constrained semi-major axis. The expected parameter degeneracies without more stringent prior constraints are also evident.

\begin{figure}
\centering
\subfigure{\includegraphics[width=0.9\columnwidth]{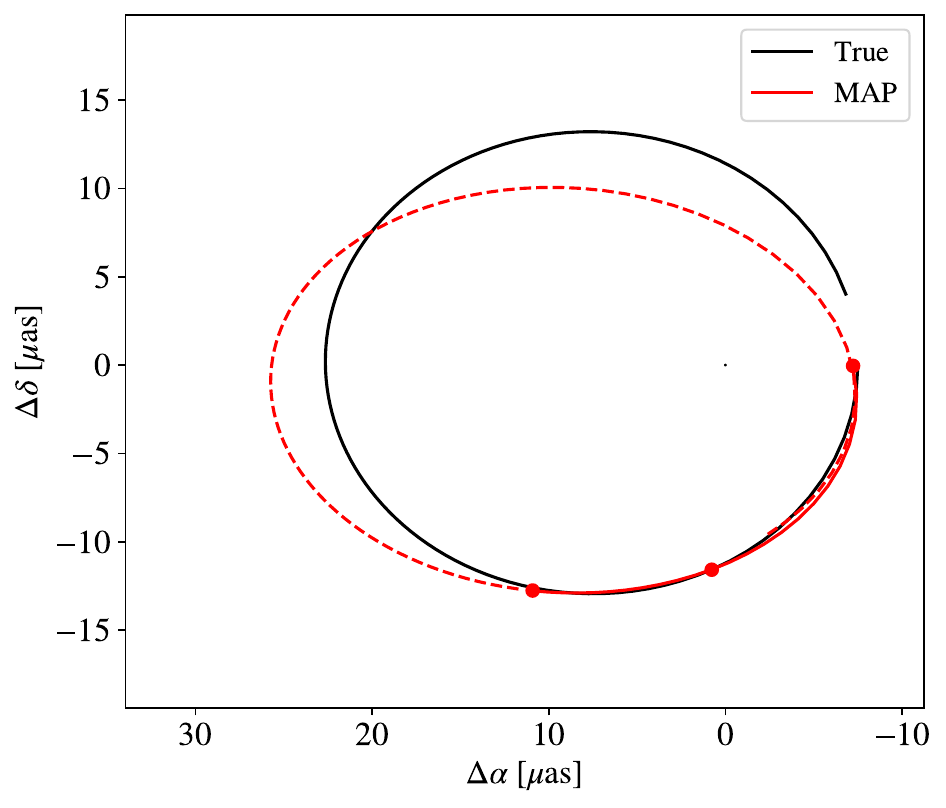}} 
\caption{Best-fit solution for Case 2.}
     \label{f:case2_bestfit}
\end{figure}

\begin{table}
\caption[]{Case 2 orbit fitting summary.}
\centering
\begin{tabular}{c c c c}
\hline
Parameter & True & MAP & Median \\ 
\hline
$M$ [$10^9$~\(\textup{M}_\odot\)] 
& 0.337
& 0.605 (79.2\%) 
& $0.561^{+0.391}_{-0.190}$ \\

$q$ [-] 
& 1.000 
& 0.898 (10.2\%) 
& $0.489^{+0.214}_{-0.338}$ \\

$a$ [pc] 
& 0.0152 
& 0.0170 (12.3\%) 
& $0.0175^{+0.1045}_{-0.00311}$ \\

$e$ [-] 
& 0.500 
& 0.566 (13.2\%) 
& $0.608^{+0.319}_{-0.109}$ \\

$i$ [$^\circ$] 
& 0 
& 38.05 
& $41.9^{+8.57}_{-32.1}$ \\

$\omega$ [$^\circ$] 
& 0 
& 147.1 
& $151.7^{+158.8}_{-94.4}$ \\

$\Omega$ [$^\circ$] 
& 0 
& 196.7 
& $215.8^{+116.4}_{-184.9}$ \\

$\tau$ [-] 
& 0 
& 0.992 
& $0.0102^{+0.980}_{-0.00872}$ \\
\hline
\label{t:case2_summary}
\end{tabular}
\end{table}

\subsubsection{Case 3: Eccentric, inclined}
\label{sss:case3}

Table \ref{t:test_summary} and \ref{t:case3_summary} summarise the model and orbit fitting results for the third test case. Fig. \ref{f:case3_bestfit} depicts the best-fit orbit. The posterior distribution corner plot is provided in Appendix \ref{As:supp}.

In general, similar trends can be seen as in Case 2, with the introduction of inclination having little impact on the posterior distribution. Semi-major axis ($a$) and eccentricity ($e$) are still the best constrained parameters. Inclination ($i$) is showing a clear tendency to a moderate inclination, but the posterior is wide. As with Case 2, a strong preference for $\tau=0/1$ is illustrated. In the MAP solution, the effect of periastron precession is also evident supporting the need for a PN propagation in the orbit fitting methodology. The effect described for Case 1, whereby the observed point near $\Delta\alpha \sim 0$ exhibits increased astrometric uncertainty, can also be seen.

Case 3 shows the effect of sampling a short arc of the orbit as beyond the three observations, the best fit orbit deviates significantly from the truth and this appears to have been exaggerated by the introduction of inclination. Additional observations would evidently help with this issue.

\begin{figure}
\centering
\subfigure{\includegraphics[width=\columnwidth]{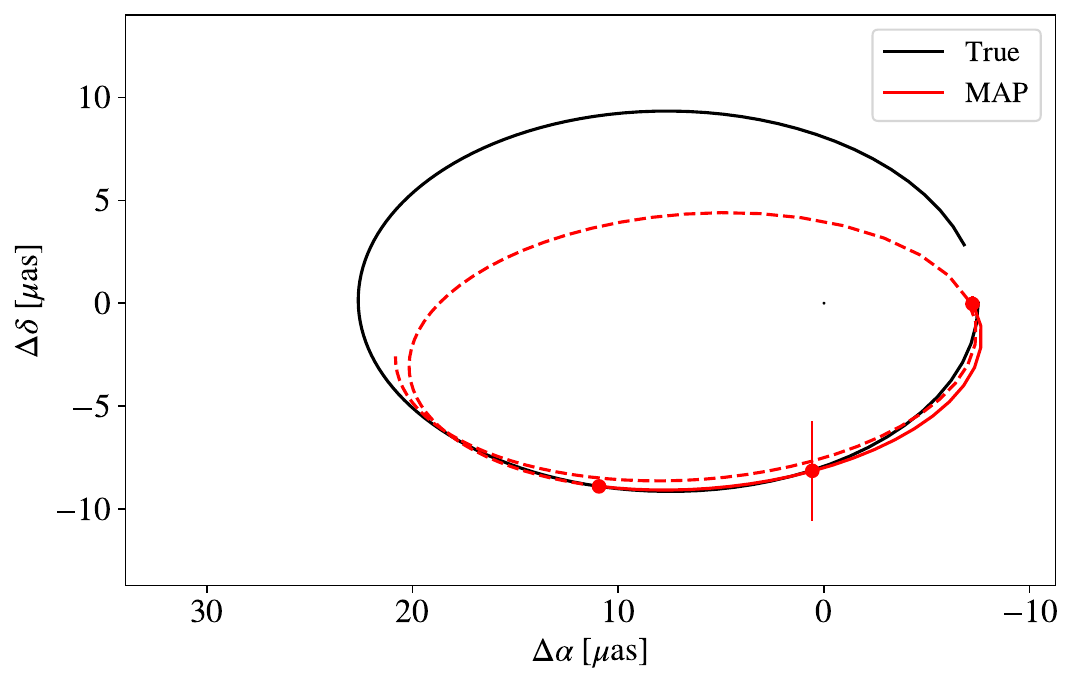}} 
\caption{Best-fit solution for Case 3.}
     \label{f:case3_bestfit}
\end{figure}

\begin{table}
\caption[]{Case 3 orbit fitting summary.}
\centering
\begin{tabular}{c c c c}
\hline
Parameter & True & MAP & Median \\ 
\hline
$M$ [$10^9$~\(\textup{M}_\odot\)] 
& 0.337 
& 0.852 (152.6\%) 
& $0.723^{+0.710}_{-0.295}$ \\

$q$ [-] 
& 1.000 
& 0.765 (23.5\%) 
& $0.494^{+0.390}_{-0.391}$ \\

$a$ [pc] 
& 0.0152 
& 0.0147 (3.09\%) 
& $0.0137^{+0.00623}_{-0.00183}$ \\

$e$ [-] 
& 0.500 
& 0.537 (7.40\%) 
& $0.522^{+0.153}_{-0.152}$ \\

$i$ [$^\circ$] 
& 45.0 
& 59.6 (32.4\%) 
& $46.6^{+19.7}_{-23.4}$ \\

$\omega$ [$^\circ$] 
& 0 
& 138.6 
& $267.5^{+47.2}_{-137.2}$ \\

$\Omega$ [$^\circ$] 
& 0 
& 182.0 
& $178.6^{+77.8}_{-167.8}$ \\

$\tau$ [-] 
& 0 
& 0.972 
& $0.958^{+0.0295}_{-0.942}$ \\
\hline
\label{t:case3_summary}
\end{tabular}
\end{table}

\subsubsection{Additional cases}
\label{sss:add}

\noindent The following cases were also tested with the synthetic data simulation and orbit fitting pipeline:

\begin{enumerate}
    \setcounter{enumi}{3}
    \item The estimated parameters of the source `Gondor' from \cite{agarwal_nanograv_2026}. Demonstration of observing a source with a wider separation at redshift $z=0.07$, with increased noise (10\% + 20~mJy);
    \item An extreme mass ratio example with total flux density similar to that of M87*, at the edge of BHEX binary detectability. Redshift $z=0.05$ and black hole masses are calculated to meet prescribed period and separation;
    \item An extreme inclination example such that the source is almost completely edge-on. Parameters the same as Case 3 but with an inclination of 80\(\degree\). This case also demonstrates the scenario of the two sources having a separation smaller than BHEX resolution;
    \item A long period example. In order to keep this case close to the fundamental millimetre plane, the separation is very wide \citep{ruffa_fundamental_2023}. As such, if it existed, such a system would likely be observable from the ground. It is included here to demonstrate the effect of a long period orbit on the orbit fitting. Redshift $z=0.05$.
\end{enumerate}

\noindent Table \ref{t:add_sims} summarises the key properties and test results for the cases defined above. The additional test cases support the trends described above whilst illustrating the effect of different parameter variations. 

Case 4 was run with an increased error budget (10\% + 20~mJy). The error bars grow noticeably as is to be expected, resulting in a significant FAR, despite the large angular separation used in the example. As the estimated source positions are still close to the actual trajectory, the MAP solutions for $a$ and $e$ are within 10\% of the true values. Case 5 depicts a scenario with a more diffuse source. The total flux density is still within the predicted detectability of BHEX as shown in Fig. \ref{f:bhex_binaries}, and therefore high relative astrometic accuracy is still achieved. The goodness-of-fit of the MAP reflects this. In Case 6, an extreme inclination example results in growth of the errors on $a$ and $e$. Interestingly, although the posterior distribution is wide, $i$ has been constrained quite well, with a clear preference for an edge-on orbit. However, the significant astrometric uncertainty on the observed point at a small angular separation to the primary causes a high FAR of 80.5\%. In general, it will be harder to confirm non-linear motion in a highly inclined system. In Case 7, a longer period case is considered of 30~years. Here, a small arc of the orbit is sampled by three BHEX observations and this is reflected in the FAR which is 100\%. The MAP solution is poorly constrained, with all parameters exhibiting wide posteriors.

\subsection{Identifying candidates}
\label{ss:cand}

Near-future capabilities will further progress the ability to identify and observe candidate binary systems. Through detection of quasi-periodic oscillation (QPO) sources, the Vera C. Rubin observatory may be able to detect hundreds to thousands of candidate SMBHBs, although with a preference for systems with short orbital periods such that statistically significant variability can be observed across the 10 year survey \citep{liao_discovery_2020}. Although certainty of a black hole binary will not be achieved with these observations, identification of likely candidates, with follow up radio observations, may provide sufficient evidence to warrant a VLBI observation. The detection of astrometric oscillations with VLBI proposed by \cite{gurvits_milliarcsecond_2025} also has potential for identification of more candidate systems.

\cite{agarwal_nanograv_2026} present a search for GW sources from 114 active galactic nuclei, using the NANOGrav 15~yr dataset, by fixing priors for period and position on the sky from QPO data and fitting the pulsar timing array measurements for the remaining binary parameters. Bayesian model comparison with uncorrelated red noise identifies eight candidates with Bayes factors (BF) $>1$. Despite the lack of statistical significance, the method by which these candidates were identified is promising and should result in more identifications, and with greater confidence in the future, as pulsar timing improves \citep{agarwal_nanograv_2026}.

As discussed in Sect. \ref{s:conditions}, \cite{valtonen_identifying_2025} demonstrate that the optical light curves and radio jet observations of OJ287 can be explained with the presence of a secondary jet. OJ287 is a target source of BHEX and the improvement in resolution by a factor of two will enable further constraining of the possible explanations for its flaring behaviour. If a secondary jet can be confidently identified, the positions of the black holes at their bases could be estimated and the method presented in this paper used to fit an orbit.

Within the ETHER database, a library of SMBHB candidates identified using the methods described previously is also being built \citep{ramakrishnan_event_2023}. Through follow-up observations of promising candidates with VLBI planned to constrain flux densities and determine detectability, the ETHER database will no doubt play an important role in identifying binary targets.


\section{Discussion}
\label{s:discuss}

The binary parameter space detectable by BHEX has been constrained, requiring $F_{\nu,\text{tot}} \geq 40$~mJy and $f\geq 0.75$ (at the minimum detectable flux density), in the thermal-noise limited case. If the secondary source is more active than the primary (as predicted in GRMHD), or the total flux of the source approaches 0.5~Jy, BHEX may be able to detect very low mass ratios ($q\lesssim0.05$). Binaries are detectable (under the conditions described in Sect. \ref{ss:bhex_detections}) even for separations down to 2~$\mu$as, albeit with increasing relative astrometric uncertainty.

As shown in Fig. \ref{f:Min_N} and Table \ref{t:error_survey}, the  relative astrometric uncertainty of BHEX observations is very low. This results in confident detection of non-linear motion for longer period orbits than is perhaps to be initially expected considering BHEX's short mission duration. Curved trajectory motion could be confidently detected in binaries with $5<a\,[\mu\text{as}]<15 $ and $7<P_\text{obs}\,\text{[years]}<28$, with only 3 annual observations. 

Across the three primary test cases, clear trends in the orbit fitting can be seen. Semi-major axis ($a$) and eccentricity ($e$) are tightly constrained to within 0.06 dex of their true values, for all cases. Total mass ($M$) exhibits a wide posterior distribution, with the MAP within 0.4 dex of the true value. Inclination ($i$) is more difficult to constrain as various combinations of $a, e$ and $i$ can produce similar projected trajectories on the sky. The position angle of nodes ($\Omega$) and argument of periastron ($\omega$) show the expected 180\(\degree\) reflection degeneracy. Mass ratio ($q$) is essentially unconstrained across all cases as, considering the Keplerian equation for orbital period, $P_\text{obs} \propto \frac{a^3}{M}$, it is only dependent on $M$ and $a$. The small number of observations that are likely to be possible with BHEX does present issues in tightly constraining orbits for which only a short arc is sampled. Naturally, more observations will always result in better fits.

The additional test cases presented in Sect. \ref{sss:add} show the effect of other parameter variations on the orbit fitting process. The limits of confident orbital motion detection are demonstrated with the long period orbit in Case 7. The challenge of inclination fitting is evident throughout the test cases, and particularly in Case 6 for an extreme, near edge-on example. Measurements of variation in the observed flux density of the source, caused by periodic Doppler boosting, could provide information on the direction of motion of the source at a given time and thus constrain the inclination \citep{dorazio_relativistic_2015}. The potential difficulty in using such a constraint is identifying Doppler boosted emission from intrinsic variability of the source.

These results hold under the assumption of 10\% fractional and 10~mJy additive noise, on a toy binary model consisting of two Gaussians. Increasing the error diminishes the ability of BHEX and the ground array to detect non-linear relative motion of the source, and to fit orbital elements. This was demonstrated with Case 4.

Table \ref{t:error_survey} shows the benefit of BHEX in binary observations. The fine angular resolution provides low relative astrometric uncertainties, increasing the orbit fitting ability of the array and potentially opening up a larger population of observable binaries. The relative astrometric accuracy of the ngEHT alone has also been evaluated, and the constraint on total number of observations disappears with ground observations. The effect of increased sampling of the orbit period can clearly be seen in the goodness of fit of the orbits for Case 1 and Case 4. The challenge, as discussed in Sect. \ref{ss:cand}, is finding candidate sources, and we acknowledge the remaining uncertainty in the sub-millimetre detectability of sub-parsec binaries.

If BHEX could successfully detect a binary source, and observe orbital motion over the course of the mission, this would be the first direct evidence of the existence of sub-parsec SMBHBs, supporting existing evidence from GW detections. Although this in itself would be a major achievement, the real scientific value would be to do this for a statistically significant population of binaries. As described by \cite{dorazio_repeated_2018}, a sample of massive binary separations and their redshift distribution would enable models for residence times to be ruled out and the parameters defining these models to be constrained. Such observations would also be beneficial in calibrating existing methods for binary candidate identification that are perhaps easier to perform than spaceborne VLBI. Such a survey would require a future spaceborne VLBI system, with a longer mission lifetime than BHEX and more specifically designed for SMBHB observation. This is discussed further in Sect. \ref{ss:prospects}.

Any observing time with BHEX dedicated to binary candidates is going to be inherently limited by the short lifetime and the primary science objectives of the mission. In this case, if sufficient observations of a candidate cannot be performed to confirm orbital motion, BHEX measurements could form part of a longer term, multi-instrument system observing the same source. This could consist of the ngEHT and/or future spaceborne VLBI systems.

\subsection{Image reconstruction}
\label{ss:imaging}

Reconstructed images from the VLBI data are the ultimate goal in the effort to directly observe an SMBHB. Generating images has multiple benefits compared to working directly with the correlated visibility data. The toy model used here is a simple depiction of a binary appearance, consisting of only two Gaussian sources with some additive noise. In reality, the binary is likely to consist of a combination of extended and evolving core emission and possibly jet launching. Reconstructing an image removes some of this complication, allowing the various emission regions to be more intuitively analysed. Also, where the \emph{(u,v)} coverage is sparse, imaging algorithms incorporate prior information to fill missing Fourier components, aiming to give a more faithful representation of the source structure. \cite{chael_interferometric_2018} provide a review of VLBI image reconstruction and present the methodology used in the \texttt{eht-imaging} package.

High fidelity image reconstruction requires as dense \emph{(u,v)} coverage as possible. Images of new black hole shadows, jets and the photon rings of M87* and Sgr\,A* are intended to be reconstructed using BHEX data. With an extensive ground array, BHEX will therefore achieve sufficient \emph{(u,v)} coverage for image reconstruction of complex sources. In this work we have shown that BHEX is capable of confidently distinguishing a binary Gaussian model from a single Gaussian. A future spaceborne VLBI mission, capable of imaging SMBHBs must be designed with the need for very dense \emph{(u,v)} coverage in mind.

\subsection{Requirements for future spaceborne VLBI}
\label{ss:prospects}

The angular resolution and sensitivity of BHEX have been primarily driven by the photon ring science. Furthermore, as shown by Fig. \ref{f:pop}, BHEX's performance is predicted to only enable observation of a few tens of binary systems, with significant uncertainty existing in this model. To observe a statistically significant number of binary systems, enough to constrain evolutionary models and draw conclusions about the distribution of binaries across redshift and in relation to their host galaxy, a more capable spaceborne interferometer is required.

As shown in Fig. \ref{f:pop}, to significantly increase the number of observable systems, angular resolution and sensitivity ideally need to be improved by an order of magnitude compared to BHEX. We set the angular resolution requirement to that originally defined for the THEZA concept, 1~\(\mu\)as. Furthermore, as shown by the results of \cite{dorazio_repeated_2018}, the number of observable systems becomes sensitivity-limited beyond this point, with minimal increase with further improving resolution. 

As described above, dense \emph{(u,v)} coverage is required for image reconstruction, with sampling of the Fourier domain of the source across a range of baseline lengths and orientations preferable. For a spaceborne system in orbit around the Earth and observing with a ground array, this is at odds with the need for fine angular resolution. The latter requires extreme distance from the Earth (with observing frequency limited by the ground array), whilst the former is more easily achieved with rapid orbital motion associated with short orbital periods at lower altitudes. To achieve both of these effects, multiple space elements are required. A proposed orbit for a dual-element THEZA system in orbit around the Earth is shown by \cite{hudson_orbital_2023} which provides the desired \emph{(u,v)} coverage features.

Scheduling observations with an extensive ground array poses considerable difficulty and limits the available observing time for a spaceborne VLBI mission. The observing frequency of the array is also effectively limited by atmospheric absorption. It is therefore desirable to have a multi-element spaceborne VLBI system, capable of observing independently from the ground to overcome these issues. As such, a minimum of three elements would enable closure phases to be measured around the triangle of antennae which provides considerable benefits for calibration and imaging \citep{chael_interferometric_2018}. Observations with a ground array should also be possible for extremely dense \emph{(u,v)} coverage and to make use of the low SEFD sites on the ground. 

Such a mission would be of a different class to BHEX, requiring a significantly higher budget and additional technological advances. \cite{gurvits_science_2022} discuss the key engineering challenges that must be tackled to realise future spaceborne VLBI missions.


\section{Conclusions}
\label{s:concl}

Obtaining conclusive electromagnetic evidence of a sub-parsec SMBHB would be a major achievement given its relevance to some of the most important ongoing research in astrophysics. Very Long Baseline Interferometry is the only astronomical technique with the resolution and sensitivity needed to directly observe such SMBHBs, and spaceborne VLBI is required to observe a statistically significant population. The BHEX mission is the most likely to be realised in the near future, and we have shown that it would theoretically be capable of providing the first detection of an SMBHB, although doing so is on the edge of its proposed capability. 

Confirming that an observed source is indeed a binary is a challenge, and observational signatures have been discussed that could be used to build confidence in a detection. Observing orbital motion with observations over multiple epochs would be the key evidence, although BHEX is not best suited to this activity given its short (2~years) mission lifetime. However, the fine resolution and extensive \emph{(u,v)} coverage provided by BHEX would be highly beneficial for reducing relative astrometric uncertainties and would perhaps enable observation of systems that would not be possible with just the ground array. The next steps to further the case for performing binary observations with BHEX requires identifying candidate sources in time for launch, and we have reviewed promising methods for this activity.

Although BHEX will be a transformational interferometer, performing a survey of a statistically significant sample of SMBHBs requires a more capable system. Estimating orbital parameters of a larger number of SMBHBs is necessary for constraining evolutionary models and drawing conclusions about their relationship with the host galaxy. Preliminary requirements of such a system have been defined as a first step towards realising a mission capable of reconstructing images of a sample of binaries. The orbit fitting methodology presented here and used throughout the paper can serve as a building block of an analysis pipeline for processing VLBI data of SMBHB candidates in the future.

\begin{acknowledgements}

This research is funded in part by the Gordon and Betty Moore Foundation, Grant GBMF12987. The authors would like to acknowledge the use of the DelftBlue computing cluster, managed by TU Delft, which was essential for running the computationally expensive examples presented in this paper \citep{DHPC2024}. We also wish to acknowledge the entirety of the BHEX community for their ongoing efforts to realise this exciting mission. BH would like to thank his employer, KISPE, for providing him with the flexibility and support to study for a PhD alongside his full-time role. Finally, the authors express their gratitude to an anonymous reviewer for their constructive comments.
\end{acknowledgements}

%
%
\bibliographystyle{aa} 
\bibliography{bibliography}


\begin{appendix}

\section{Orbit model error analysis}
\label{A:orbit_orders}

The PN orbit propagation used to evaluate the likelihood of candidates in the orbit fitting method has been implemented up to order 3.5PN. Provided in this section is an evaluation of the order of magnitude impact of higher orders and other relativistic effects on the observed black hole positions.

The terms up to 3PN produce the characteristic precession of an eccentric orbit that occurs on the order of single orbital periods. This behaviour is desired in this model as the precessing nature of the orbit considerably changes the observed position within the timescales of possible VLBI observations. The candidate SMBHB OJ287 exhibits \(\sim\)40\(\degree\) per \(\sim\)12 year orbit. 

Each PN term has an associated expansion factor $(\frac{v}{c})^{2PN}$ where $v$ is the orbital velocity vector. The PN terms therefore have the greatest effect when $v$ is a maximum. Consider an orbit with $v=0.1c$, over an orbital period of 20 years. Approximating the differential displacement due to 4PN acceleration term as $\Delta x \sim \frac{1}{2} \Delta a_N t^2$ where $a_N$ is the Newtonian acceleration given by $\frac{GM}{r^2}$, a source at redshift 0.2 will only exhibit an angular displacement of $10^{-6}$ \(\mu\)as, far below resolution limits of VLBI.

The spin-orbit coupling effect produces a precession of the orbital plane which can be calculated from the equation provided by \cite{dey_authenticating_2018}. Considering the same example orbit as above, and black hole spins of $\chi_1 = \chi_2 = 0.5$, the precession rate after a single orbital period equates to a differential angular displacement of \(\sim\)0.015~\(\mu\)as. The spin-spin coupling effect causes a precession of a lower order that results in an angular displacement of \(\sim\)$10^{-4}$~\(\mu\)as.

For an inclined binary orbit, the relative time for light to travel from each source to the observer results in the farther black hole appearing in a position from an earlier time than the closer one. Considering the orbit above with a worst-case inclination of 90\(\degree\), this effect causes the secondary black hole to appear at a position \(\sim\)0.6~\(\mu\)as different to its real location at the time of the primary black hole's emission. At a tenth of the observed separation, this is by far the largest effect compared to higher orders of the PN propagation. With BHEX achieving astrometric uncertainties of \(\sim\)0.1~\(\mu\)as under certain noise conditions, this is non-negligible. Table \ref{t:orbit_errors} summarises these results.

To include the relative light travel time effect, the emission time $t_{emit}$ of a black hole (index $i$) observed at time $t_{obs}$ in the binary rest frame is calculated from

\begin{equation}
    t_{\text{emit,i}} = t_{\text{obs,i}} - \frac{1+z}{c} D_{\text{los}}(t_{\text{emit,i}}),
\end{equation}

\noindent where $D_{\text{los}}(t_{\text{emit,i}}) = \mathbf{x}_{i}(t_{\text{emit,i}})\cdot\mathbf{\hat{n}}$ is the distance along the observer line of sight of the black hole from the binary barycentre. $\hat{\textbf{n}}$ is the observer unit vector direction, which in this case is the z-axis. This is solved iteratively using a simple numerical method

\begin{equation}
    t_{\text{emit,i}}^{n+1} = t_{\text{obs,i}} - \frac{1+z}{c} D_{\text{los}}(t_{\text{emit,i}}^{n}).
\end{equation}

\noindent The observed position of each black hole is then $\mathbf{x}_{i}^{\text{obs}}(t_{\text{obs}}) = \mathbf{x}_{i}(t_{\text{emit,i}})$. The coordinate system is transformed to place the primary black hole at the origin and the relative right ascension and declination of the secondary are computed by projecting the observed position in the binary rest frame onto the sky plane and converting to angular units. This approach also includes the time dilation effect of the source being at redshift $z$ on the relative light travel time.

\renewcommand{\arraystretch}{1.1}
\begin{table}
\caption[]{Orbit model error analysis.}
\centering 
\begin{tabular}{c c}
\hline
Effect & $\sim\Delta \theta$ [\(\mu\)as] \\
\hline
Relative light travel time & 0.6 \\
Spin-orbit coupling & 0.015 \\
Spin-spin coupling & $10^{-4}$ \\
$\geq$4PN terms & $10^{-6}$ \\
\hline
\label{t:orbit_errors}
\end{tabular}
\tablefoot{Differential angular displacements provided for an example orbit with $P_{\text{obs}}$ = 20~years, $a = 6~\mu$as, $v = 0.1c$.}
\end{table}

\section{Minimum number of observations}
\label{A:linear_motion}

In this section is provided the full derivation of Eq. \ref{E:min_n}. For a face-on, circular orbit, the projected motion on the sky in  right ascension $\alpha$ is
\[
\alpha = a \cos\left( \frac{2\pi t}{P_{\text{obs}}} \right),
\]

\noindent where $a$ is the orbit semi-major axis in \(\mu\)as. For small \( t \ll P_{\text{obs}} \), we expand in a Taylor series,

\[
\alpha \approx a - \frac{1}{2} a \left( \frac{2\pi t}{P_{\text{obs}}} \right)^2 = a \left( 1 - \frac{1}{2} \left( \frac{2\pi t}{P_{\text{obs}}} \right)^2 \right).
\]

\noindent Thus, the maximum deviation from linear motion over a time span \( T \) is approximately
\[
\Delta \alpha_{\text{max}} \sim a \left( \frac{\pi T}{P_{\text{obs}}} \right)^2.
\]

\noindent Assuming \( N_{min} \) observations spaced by cadence \( \tau \), the total span is \( T = (N_{min} - 1)t_{\text{cad}} \), and therefore
\[
\Delta \alpha_{\text{max}} \sim a \left( \frac{\pi (N_{min} - 1) t_{\text{cad}}}{P_{\text{obs}}} \right)^2.
\]

\noindent To statistically detect this curvature, we require
\[
\Delta \alpha_{\text{max}} \gtrsim k \sigma_{\text{pos}},
\]

\noindent for some significance threshold \( k \) (e.g. \( k = 3 \) or \( 5 \)) and where $\sigma_{\text{pos}}$ is the uncertainty in the position estimation of the secondary black hole. Substituting and solving for \( N_{\min} \), we get
\[
a \left( \frac{\pi (N_{min} - 1) t_{\text{cad}}}{P_{\text{obs}}} \right)^2 \geq k \sigma_{\text{pos}},
\]

\[
N_{min} \geq 1 + \sqrt{ \frac{k \varepsilon P_{\text{obs}}^2}{a \pi^2 t_{\text{cad}}^2} }.
\]

\noindent Since at least three observations are required to geometrically detect curvature, we enforce

\[
N_{min} = \max\left( 3,\ \Biggl \lceil 1 + \sqrt{ \frac{k \sigma_{\text{pos}} P_{\text{obs}}^2}{a \pi^2 t_{\text{cad}}^2 \cos{i}} } \Biggr\rceil \right).
\]

\noindent The effect of inclination can be introduced with the simple addition of a $\cos{i}$ term in the denominator as it reduces the projected separation by this factor.

The derivation of this equation approximates the curvature of a circular orbit using a Taylor expansion to the quadratic order. As such, its accuracy diminishes as the total elapsed time of observations approaches half of the orbital period of the binary. However, this will always underestimate the actual curvature of the orbit, providing a conservative estimate of $N_{\text{min}}$. 

\section{Simulated observations}
\label{A:sim_obs}

\subsection{Simulation configuration}
\label{As:sim_config}

Provided in this section are the definitions of key simulation properties used in the \texttt{ngehtsim} and \texttt{eht-imaging} simulations presented in Sect. \ref{ss:bhex_detections}. Table \ref{t:sim_conf} describes the modelled parameters for BHEX. Table \ref{t:ground_conf} defines the ground array used in the simulation. Provided for ALMA is an effective diameter that accounts for multiple antenna being phased together during observations. The reader is referred to the full documentation of \texttt{eht-imaging} and \texttt{ngehtsim} for descriptions of how these simulations are performed.

Weather conditions at each ground site are also modelled, and an assumption of median weather conditions is used for these simulations.  BHEX will utilise a northern ground array for January--March observations and a southern array for June--August. For all cases in Sect. \ref{ss:bhex_orbital}, it is assumed the source lies close to M87*, and therefore the January--March window is utilised, with ALMA participating in the observations but not performing FPT. We fit amplitudes and closure phases due to the absence of phase calibration in typical VLBI.

In our previous work, the various functional constraints impacting a space-based VLBI mission were analysed with methods for mitigation proposed for BHEX's specific challenges \citep{hudson_toward_2025}. Apart from Sun and Earth blinding of the main antenna, these effects are not included in these simulations due to lack of engineering maturity of the concept. BHEX will observe with a cadence of every 1 hour, resulting in full azimuthal sampling of the \emph{(u,v)} plane, across its \(\sim\)12~hour orbit.

\begin{table}
\caption[]{BHEX properties.}
\centering 
\begin{tabular}{c c}
\hline
Parameter & Value \\ 
\hline
    Antenna Diameter & 3.4~m \\ 
    Surface Accuracy & 40~$\mu$m \\
    High-band LO Frequency & 308~GHz \\ 
    Low-band LO Frequency & 86~GHz\\
    Receiver Noise Temperature & 40~K\\ 
    Bandwidth & 8~GHz \\ 
    Pre-FPT Low-band $t_{int}$ & 45~s \\ 
    Pre-FPT High-band $t_{int}$ & 10~s  \\
    Post-FPT High-band $t_{int}$ & 600~s  \\
    Quantisation & 1 bit \\
\hline
\label{t:sim_conf}
\end{tabular}
\end{table}

\begin{table}
\caption[]{Array properties.}
\centering
\begin{tabular}{c c c c c}
\hline
Site & Array & Diameter & SEFD  & SEFD  \\
& & [m] & (High-band) & (Low-band) \\
\hline
    ALMA & N \& S & 75 & 157 -- 163 & N/A \\
    LMT & N \& S & 50 &  854 -- 876 & 169 -- 171 \\
    SMA & N \& S & 15 & 3313 -- 3491 & N/A \\
    JCMT & N \& S & 15 & 3408 -- 3591 & 1680 -- 1688 \\
    IRAM & N & 30 & 1955 -- 2177 & N/A \\
    GLT & N & 12 & 18824 -- 20008 & N/A \\
    KP & N & 12 & N/A & 3093 -- 3197\\
    GBT & N & 100 & N/A & 204 -- 208 \\
\hline
    BHEX & N \& S & 3.4 & 39817 & 19207 \\
\hline
\label{t:ground_conf}
\end{tabular}
\tablefoot{Median System Equivalent Flux Density (SEFD) ranges provided for high-band and low-band, accounting for variation across the three observing sessions. Where N/A is noted, this site cannot observe in the defined band. For each site, it is defined whether it participates as part of the Northern (N) array or Southern (S) or both.}
\end{table}

\subsection{Supplementary simulation material}
\label{As:supp}

\begin{figure*}
\centering
\includegraphics[width=0.8\textwidth]{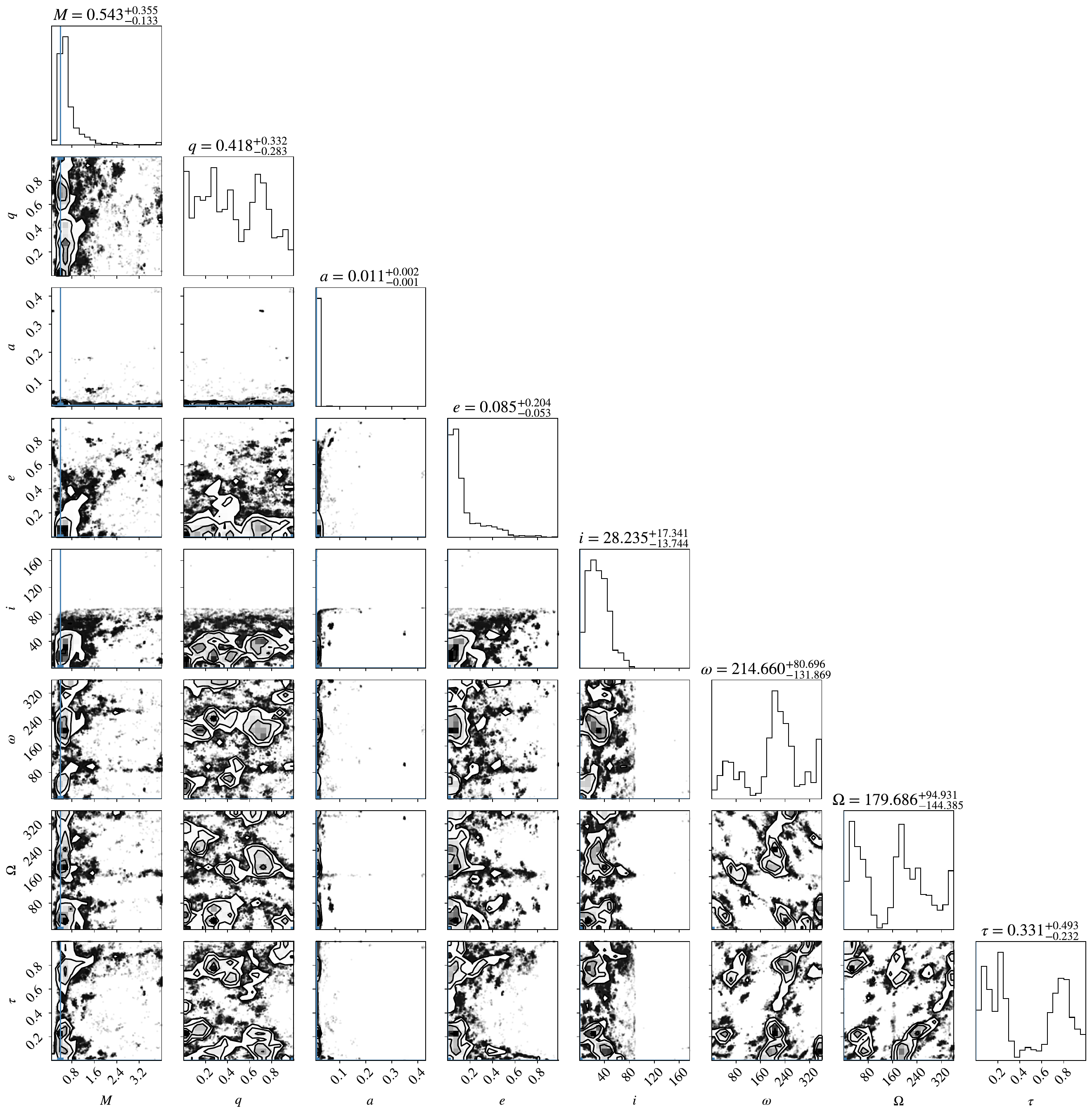}
  \caption{Case 1 corner plot.}
     \label{f:case1_corner}
\end{figure*}

\begin{figure*}
\centering
\includegraphics[width=0.8\textwidth]{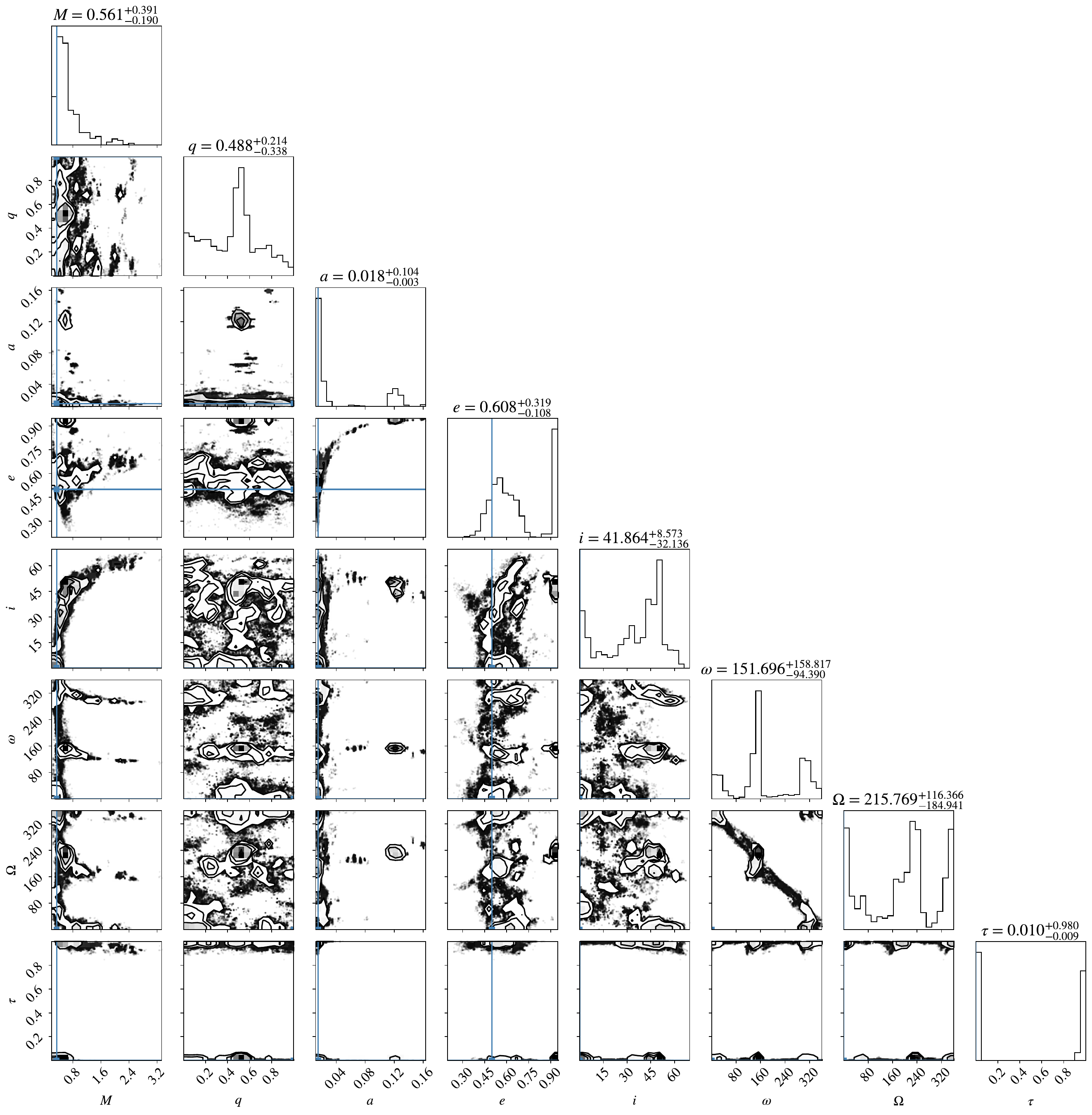}
  \caption{Case 2 corner plot.}
     \label{f:case2_corner}
\end{figure*}

\begin{figure*}
\centering
\includegraphics[width=0.8\textwidth]{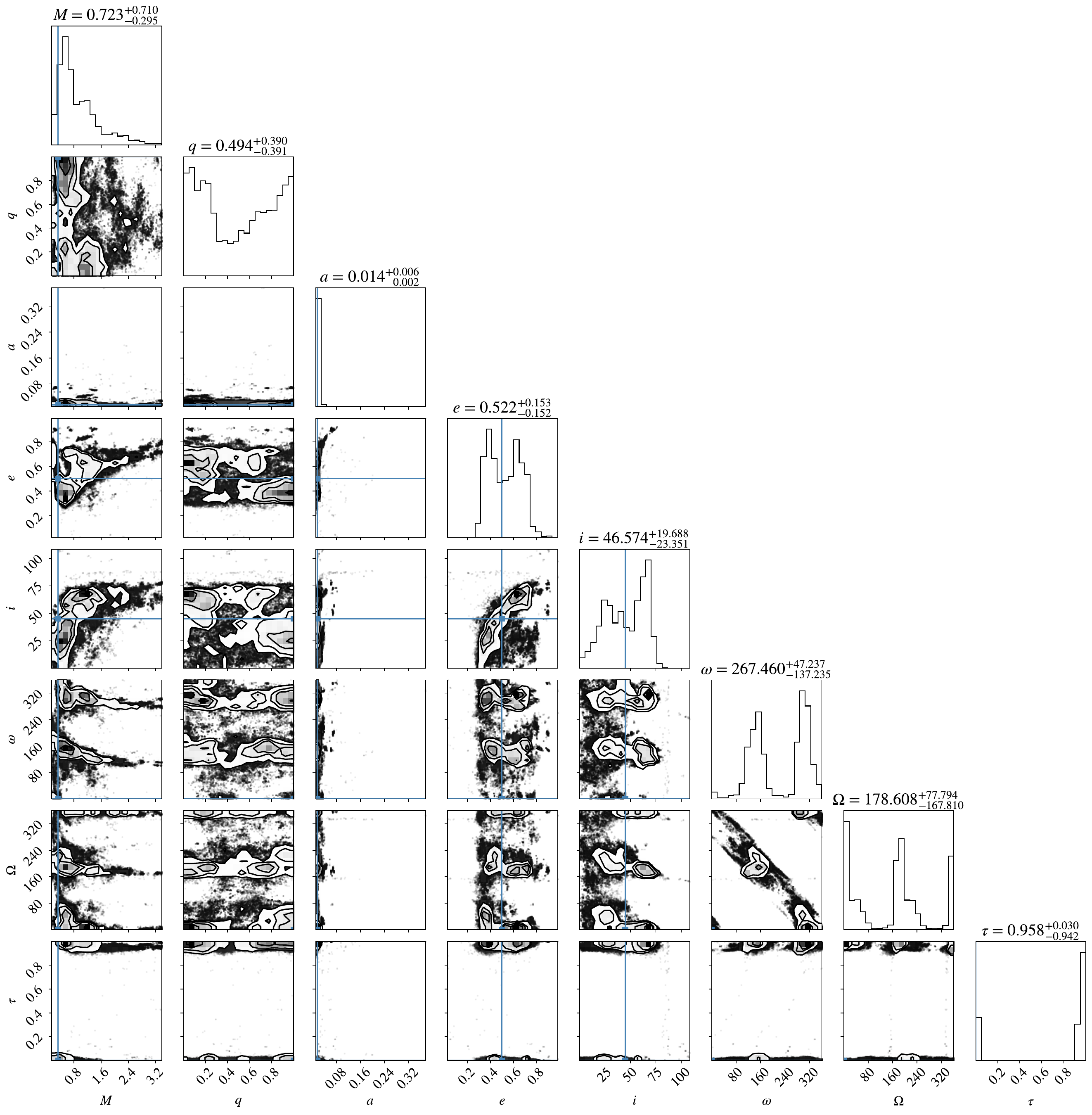}
  \caption{Case 3 corner plot.}
     \label{f:case3_corner}
\end{figure*}

\noindent Supplementary figures for Case 1, 2 and 3, and the additional test cases are provided in this section. Figures \ref{f:case1_corner}, \ref{f:case2_corner} and \ref{f:case3_corner} show the posterior distribution of the runs in the form of corner plots. Tables \ref{t:case1_supp_summary}, \ref{t:case2_supp_summary}, \ref{t:case3_supp_summary} and \ref{t:case4_supp_summary} present the orbit fitting results of the additional test cases.

\renewcommand{\arraystretch}{1.3}
\begin{table*}
\caption[]{Supplementary test case summary.}
\centering
\begin{tabular}{c c c c c c c c c c c}
\hline
Case & $F_{\nu,\text{tot}}$ [Jy] & $q$ & $P_{\text{obs}}$ [yr] & Separation [\(\mu\)as] & $e$ [-] & $i$ [$\degree$] & $M$ [$10^8$ \(\textup{M}_\odot\)] & $\text{FWHM}_{1/2}$ [\(\mu\)as]   \\
\hline
4 & 0.03/0.02 & 0.6 & 4.41 & 15 & 0 & 0 & 17.3 & 2/1.6  \\
5 & 0.47/0.03 & 0.06 & 5 & 10 & 0 & 0 & 3.37 & 5/0.3   \\
6 & 0.03/0.02 & 1 & 10 & $r_{p}=7.5$, $r_{a}=22.5$ & 0.5 & 80 & 3.37 & 2/2  \\
7 & 0.03/0.02 & 1 & 30 & 30 & 0 & 0 & 3.00 & 2/2  \\
\hline
&  $\nu L_{\nu}$ [ergs$^{-1}$] & $\chi^2_{amp}$ & $\chi^2_{cp}$ & $\chi^2_{amp,sin}$ & $\chi^2_{cp,sin}$ & $\sigma_\text{pos,av}$ [\(\mu\)as] & $\chi^2_{orb}$ & FAR [\%]\\
\hline
4 & $6.07/4.05\times10^{41}$ & 0.11 & 0.02 & 0.48 & 0.27 & 2.39 & 0.02 & 48\\
5 & $9.90/0.55\times10^{42}$ & 0.03 & 0.24 & 0.47 & 0.36 & 0.12 & 0.06 & 0 \\
6 & 5.54/3.68$\times 10^{41}$ & 0.20 & 0.21 & 7.27 & 0.38 & 0.64 & 0.02 & 81 \\
7 & 5.54/3.68$\times 10^{41}$ & 0.08 & 0.05 & 0.91 & 0.17 & 1.16 & 0.14 & 100  \\
\hline
\label{t:add_sims}
\end{tabular}
\tablefoot{For all cases, $\Omega$, $\omega$ and $\tau$ are set to 0, as they are expected to have less impact on the orbit fitting than the varied parameters. $\chi^2_{amp}$ and $\chi^2_{cp}$ describe the average goodness of fit of the binary Gaussian model in amplitude and closure phase to the observed data in \texttt{eht-imaging}, across the three observation epochs. The subscript $sin$ describes the $\chi^2$ values for fitting a single Gaussian source to the observed data. $\chi^2_{orb}$ describes the goodness of fit of the MAP orbital solution.}
\end{table*}

\renewcommand{\arraystretch}{1.3}
\begin{table}
\caption[]{Case 4 orbit fitting summary.}
\centering
\begin{tabular}{c c c c}
\hline
Parameter & True & MAP & Median \\
\hline
$M$ [$10^9$ \(\textup{M}_\odot\)] & 1.734  & 2.156 (24.3\%) & $6.904^{\num{+4.996}}_{\num{-5.085}}$ \\
$q$ [-]                          & 0.600  & 0.131 (78.2\%) & $0.722^{\num{+0.146}}_{\num{-0.526}}$ \\
$a$ [pc]                         & 0.0152 & 0.0159 (4.85\%) & $0.0167^{\num{+0.00458}}_{\num{-0.00339}}$ \\
$e$ [-]                          & 0.000  & 0.00111 & $0.276^{\num{+0.138}}_{\num{-0.177}}$ \\
$i$ [$\degree$]                  & 0.000  & 25.7 & $37.9^{\num{+51.9}}_{\num{-24.7}}$ \\
$\omega$ [$\degree$]             & 0.000  & 342.3 & $161.7^{\num{+146.5}}_{\num{-102.3}}$ \\
$\Omega$ [$\degree$]             & 0.000  & 35.1 & $176.1^{\num{+86.4}}_{\num{-120.7}}$ \\
$\tau$ [-]                       & 0.000  & 0.0492 & $0.623^{\num{+0.248}}_{\num{-0.453}}$ \\
\hline
\label{t:case1_supp_summary}
\end{tabular}
\end{table}

\begin{figure}
\centering
\subfigure{\includegraphics[width=0.9\columnwidth]{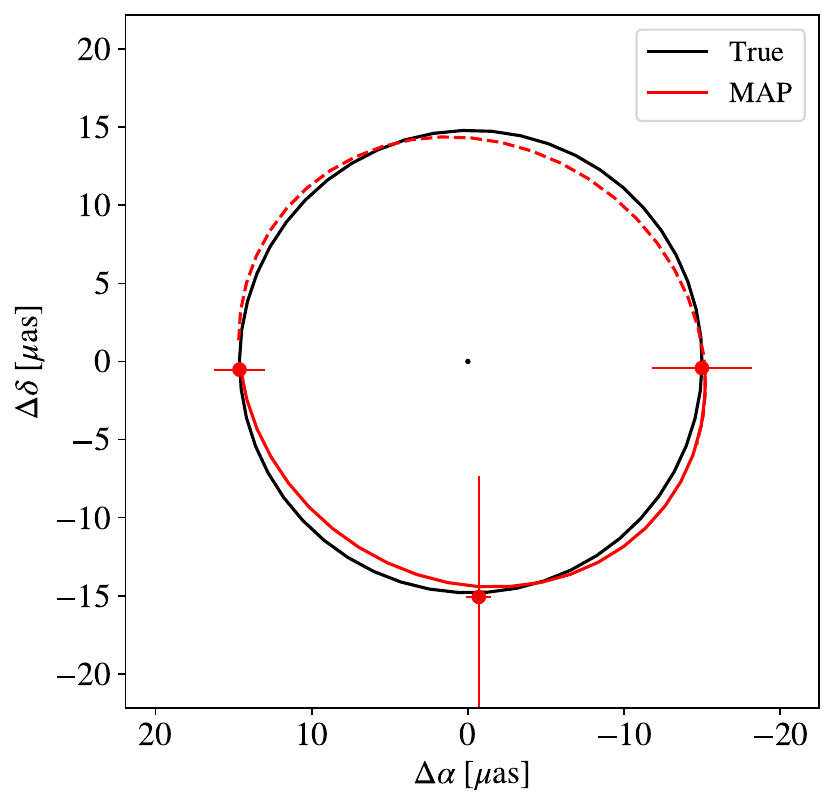}} 

  \caption{Best-fit solution for Case 4.}
     \label{f:case4_bestfit}
\end{figure}

\begin{table}
\caption[]{Case 5 orbit fitting summary.}
\centering
\begin{tabular}{c c c c}
\hline
Parameter & True & MAP & Median \\ 
\hline
$M$ [$10^9$~\(\textup{M}_\odot\)] & 5.62 & 1.296 (76.9\%) & $1.342^{+24.87}_{-0.196}$ \\
$q$ [-] & 0.600 & 0.935 (55.8\%) & $0.475^{+0.425}_{-0.354}$ \\
$a$ [pc] & 0.0138 & 0.0145 (4.9\%) & $0.0146^{+0.0036}_{-0.0005}$ \\
$e$ [-] & 0.000 & 0.031 & $0.0779^{+0.299}_{-0.0588}$ \\
$i$ [$^\circ$] & 0 & 25.5 & $24.7^{+96.2}_{-13.2}$ \\
$\omega$ [$^\circ$] & 0 & 295.9 & $225.4^{+73.1}_{-167.1}$ \\
$\Omega$ [$^\circ$] & 0 & 24.0 & $176.6^{+105.4}_{-153.7}$ \\
$\tau$ [-] & 0 & 0.901 & $0.797^{+0.109}_{-0.660}$ \\
\hline
\label{t:case2_supp_summary}
\end{tabular}
\end{table}

\begin{figure}
\centering
\subfigure{\includegraphics[width=0.9\columnwidth]{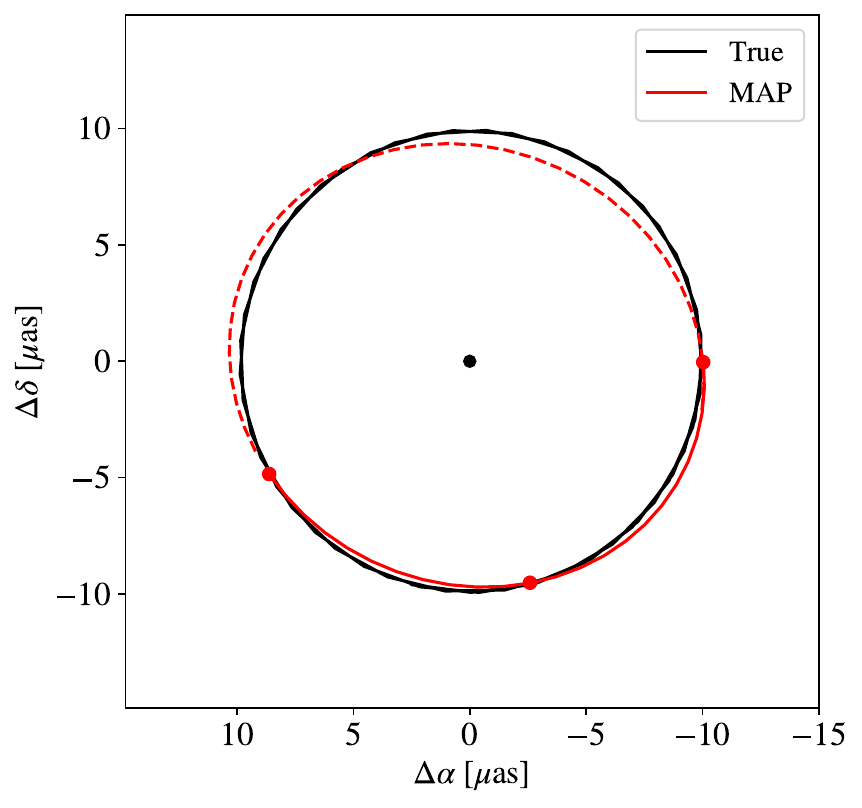}} 

  \caption{Best-fit solution for Case 5.}
     \label{f:case5_bestfit}
\end{figure}

\begin{table}
\caption[]{Case 6 orbit fitting summary.}
\centering
\begin{tabular}{c c c c}
\hline
Parameter & True & MAP & Median \\ 
\hline
$M$ [$10^9$~\(\textup{M}_\odot\)] & 0.337 & 0.776 (130\%) & $1.282^{+0.669}_{-0.927}$ \\
$q$ [-] & 1.000 & 0.124 (87.7\%) & $0.392^{+0.455}_{-0.313}$ \\
$a$ [pc] & 0.0152 & 0.0103 (31.8\%) & $0.0116^{+0.0102}_{-0.0009}$ \\
$e$ [-] & 0.500 & 0.164 (67.2\%) & $0.170^{+0.570}_{-0.112}$ \\
$i$ [$^\circ$] & 80.0 & 101.3 & $99.4^{+9.3}_{-3.1}$ \\
$\omega$ [$^\circ$] & 0 & 228.5 & $226.0^{+33.6}_{-181.2}$ \\
$\Omega$ [$^\circ$] & 0 & 169.5 & $171.9^{+178.2}_{-2.4}$ \\
$\tau$ [-] & 0 & 0.208 & $0.234^{+0.671}_{-0.221}$ \\
\hline
\label{t:case3_supp_summary}
\end{tabular}
\end{table}

\begin{figure}
\centering
\subfigure{\includegraphics[width=\columnwidth]{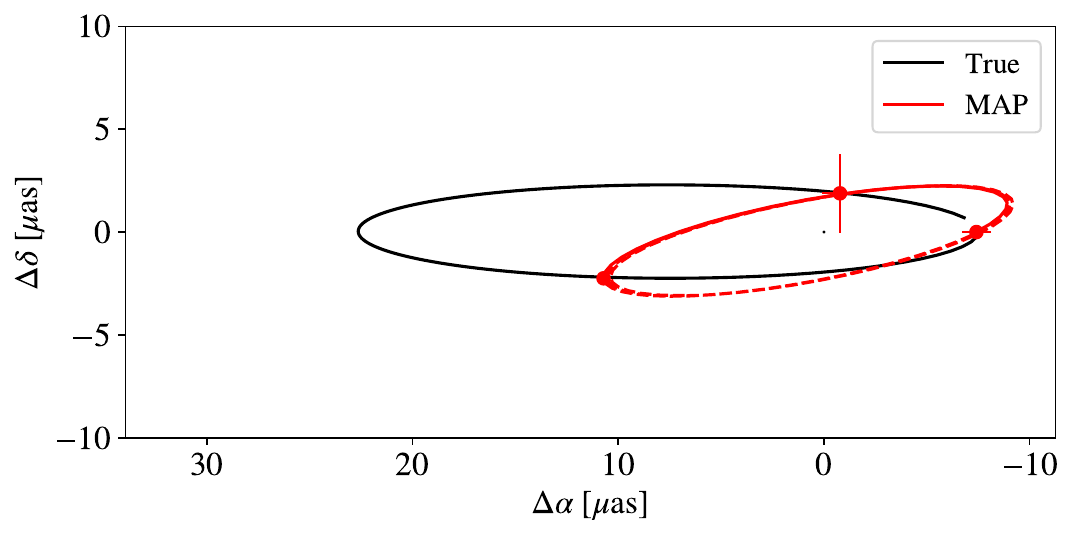}} 

  \caption{Best-fit solution for Case 6.}
     \label{f:case6_bestfit}
\end{figure}

\begin{table}
\caption[]{Case 7 orbit fitting summary.}
\centering
\begin{tabular}{c c c c}
\hline
Parameter & True & MAP & Median \\ 
\hline
$M$ [$10^9$~\(\textup{M}_\odot\)] 
& 0.300 
& 1.722 (474\%) 
& $0.808^{+1.224}_{-0.461}$ \\

$q$ [-] 
& 1.000 
& 0.817 (18.3\%) 
& $0.507^{+0.332}_{-0.353}$ \\

$a$ [pc] 
& 0.0303 
& 0.0735 (142\%) 
& $0.0429^{+0.0788}_{-0.0146}$ \\

$e$ [-] 
& 0.000 
& 0.415 
& $0.355^{+0.352}_{-0.282}$ \\

$i$ [$^\circ$] 
& 0 
& 73.3 
& $52.9^{+20.4}_{-35.6}$ \\

$\omega$ [$^\circ$] 
& 0 
& 94.8 
& $150.2^{+157.0}_{-97.9}$ \\

$\Omega$ [$^\circ$] 
& 0 
& 102.2 
& $137.3^{+160.7}_{-70.5}$ \\

$\tau$ [-] 
& 0 
& 0.550 
& $0.499^{+0.375}_{-0.408}$ \\
\hline
\label{t:case4_supp_summary}
\end{tabular}
\end{table}

\begin{figure}
\centering
\subfigure{\includegraphics[width=0.9\columnwidth]{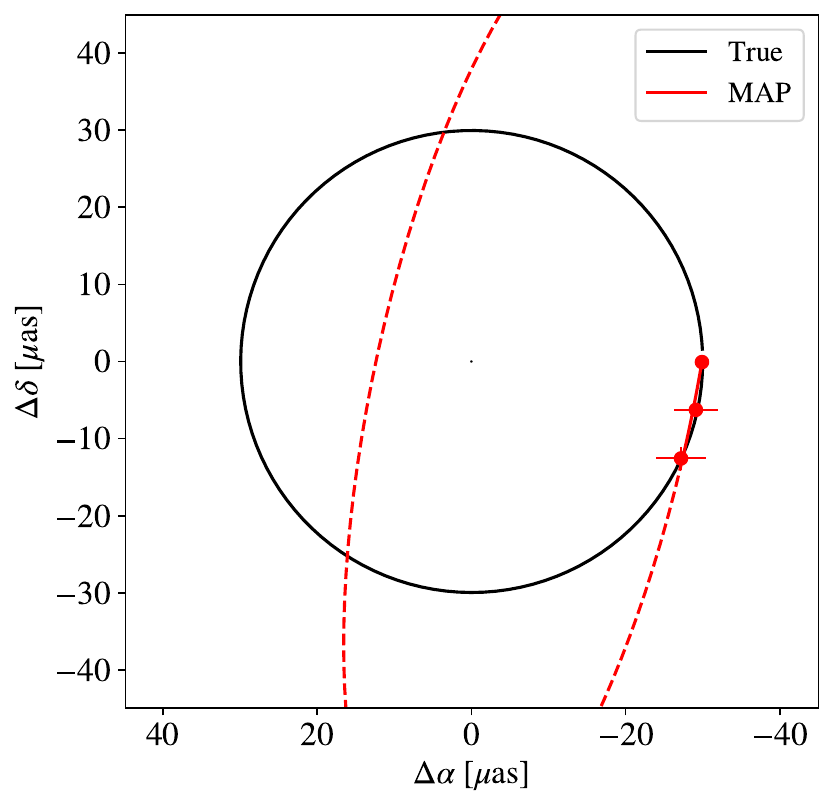}} 

  \caption{Best-fit solution for Case 7.}
     \label{f:case7_bestfit}
\end{figure}

\end{appendix}

\end{document}